# Worldwide Patterns of Ancestry, Divergence, and Admixture in Domesticated Cattle

## Short Title:
## Ancestry, Divergence, and Admixture in Cattle

Jared E. Decker, Division of Animal Sciences, University of Missouri, Columbia, Missouri, United States of America

Stephanie D. McKay, Department of Animal Science, The University of Vermont, Burlington, Vermont, United States of America

Megan M. Rolf, Department of Animal Sciences, Oklahoma State University, Stillwater, Oklahoma, United States of America

JaeWoo Kim, Division of Animal Sciences, University of Missouri, Columbia, Missouri, United States of America

Antonio Molina Alcalá, Departamento de Genética, Facultad de Veterinaria, Universidad de Córdoba, Córdoba, Spain

Tad S. Sonstegard, USDA-ARS Bovine Functional Genomics Lab, Beltsville, Maryland, United States of America

Olivier Hanotte, Medicine & Health Sciences, The University of Nottingham, Nottingham, United Kingdom

Anders Götherström, Evolutionary Biology Centre, Uppsala Universitet, Uppsala, Sweden

Christopher M. Seabury, Department of Veterinary Pathobiology, College of Veterinary Medicine, Texas A&M University, College Station, Texas, United States of America

Lisa Praharani, Indonesian Research Institute for Animal Production, Ciawi, Bogor, Indonesia

Masroor Ellahi Babar, Institute of Biochemistry & Biotechnology, University of Veterinary and Animal Sciences, Lahore, Pakistan

Luciana Correia de Almeida Regitano, Animal Molecular Genetics, Embrapa Pecuaria Sudeste, Sao Carlos - Sao Paulo, Brasil

Mehmet Ali Yildiz, Animal Science, Biometry and Genetics, Ankara University, Diskapi – Ankara, Turkey

Michael P. Heaton, USDA-ARS Meat Animal Research Center, Clay Center, Nebraska, United States of America

Wan-sheng Liu, Department of Animal Science, Pennsylvania State University, University Park, Pennsylvania, United States of America

Chu-Zhao Lei, College of Animal Science and Technology, Northwest A&F University, Yangling, Shaanxi 712100, China

James M. Reecy, Department of Animal Science, Iowa State University, Ames, Iowa, United States of America

Muhammad Saif-Ur-Rehman, Department of Animal Breeding and Genetics, University of Agriculture, Faisalabad, Pakistan

Robert D. Schnabel, Division of Animal Sciences, University of Missouri, Columbia, Missouri, United States of America

Jeremy F. Taylor, Division of Animal Sciences, University of Missouri, Columbia, Missouri, United States of America

**Corresponding Authors**
Jared E. Decker
Email: DeckerJE@missouri.edu
Jeremy F. Taylor
E-mail: Taylorjerr@missouri.edu

## Abstract

The domestication and development of cattle has considerably impacted human societies, but the histories of cattle breeds have been poorly understood especially for African, Asian, and American breeds. Using genotypes from 43,043 autosomal single nucleotide polymorphism markers scored in 1,543 animals, we evaluate the population structure of 134 domesticated bovid breeds. Regardless of the analytical method or sample subset, the three major groups of Asian indicine, Eurasian taurine, and African taurine were consistently observed. Patterns of geographic dispersal resulting from co-migration with humans and exportation are recognizable in phylogenetic networks. All analytical methods reveal patterns of hybridization which occurred after divergence. Using 19 breeds, we map the cline of indicine introgression into Africa. We infer that African taurine possess a large portion of wild African auroch ancestry, causing their divergence from Eurasian taurine. We detect exportation patterns in Asia and identify a cline of Eurasian taurine/indicine hybridization in Asia. We also identify the influence of species other than *Bos taurus* in the formation of Asian breeds. We detect the pronounced influence of Shorthorn cattle in the formation of European breeds. Iberian and Italian cattle possess introgression from African taurine. American Criollo cattle are shown to be imported from Iberia, and not directly from Africa, and African ancestry is inherited via Iberian ancestors. Indicine introgression into American cattle occurred in the Americas, and not Europe. We argue that cattle migration, movement and trading followed by admixture have been important forces in shaping modern bovine genomic variation.

## Author Summary

The DNA of domesticated plants and animals contains information about how species were domesticated, exported, and bred by early farmers. Modern breeds were developed by lengthy and complex processes; however, our use of 134 breeds and new analytical models enabled us to reveal some of the processes that created modern cattle diversity. In Asia, Africa, North and South America, humpless (*Bos t. taurus* or taurine) and humped (*Bos t. indicus* or indicine) cattle were crossbred to produce hybrids adapted to the environment and endemic production systems. The history of Asian cattle involves the domestication and admixture of several species whereas African taurines arose through the introduction of domesticated Fertile Crescent taurines and their hybridization with wild African aurochs. African taurine introgression was common among European Mediterranean breeds. The absence of indicine introgression within most European taurine breeds, but presence within three Italian breeds is consistent with two separate migrations, one from the Middle East which captured taurines in which indicine introgression had already occurred and the second from western Africa into Spain with no indicine introgression. This second group seems to have radiated from Spain into the Mediterranean resulting in a cline of African taurine introgression into European taurines.

## Introduction

High-throughput genotyping assays have allowed population geneticists to use genome-wide marker sets to analyze the histories of many species, including human [1], cattle [2–4], sheep [5], dog [6], horse [7], yeast [8], mouse [9,10], rice [11,12], maize [13–16], grape [17], and wheat [18]. We previously described the phylogeny of domesticated bovine populations using their genetic variation inferred from a sample of 40,843 single-nucleotide polymorphisms (SNPs) [3]. Although we had sampled 48 cattle breeds, we did not have samples from key geographic regions including China and Southeast Asia, Anatolia, the Baltic States, southern and eastern Africa, and the Iberian Peninsula. As a consequence of those gaps in geographic sampling, we were unable to address the origins of cattle in these regions and the extent to which these cattle influenced the population structure of regions such as the New World.

We have now assembled a genomic data set which represents the largest population sampling of any mammalian species. This allows for an extremely detailed description of the population structure of domesticated cattle worldwide. Using this data set, we accurately establish the patterns of exportation, divergence, and admixture for domesticated cattle.

## Materials and Methods

### Sample selection

We used 1,543 samples in total, including 234 samples from [3] and 425 samples from [4], see Table S1. We selected samples that had fewer than 10% missing genotypes, and for breeds with fewer than 20 genotyped samples, we used all available samples which passed the missing genotype data threshold. When pedigree data were absent for a breed, the 20 samples with the highest genotype call rates were selected. For breeds which had pedigree information, we filtered any animals whose sire or dam was also genotyped. For identified half-siblings, we sampled only the sibling with the highest genotype call rate. After removing genotyped animals known to be closely related, we selected the 20 animals with the highest genotype call rate to represent the breed.

### Genotyping

Samples were genotyped with the Illumina BovineSNP50 BeadChip [19]. Autosomal SNPs and a single pseudoautosomal SNP were analyzed, because the data set from Gautier et al. [4] excluded SNPs located exclusively on the X chromosome. We also filtered all SNPs which mapped to “chromosome unknown” of the UMD3.1 assembly [20]. In PLINK [21,22], we removed SNPs with greater than 10% missing genotypes and with minor allele frequencies less than 0.0005 (1/[2*Number of Samples] = 0.000324, thus the minor allele had to be observed at least once in our data set). The average total genotype call rate in the remaining individuals was 0.993. Genotype data were deposited at DRYAD [23].

## Principal component analysis

The sample genotype covariance matrix was decomposed using SMARTPCA, part of EIGENSOFT 4.2 [24]. To limit the effects of linkage disequilibrium on the estimation of principal components, for each SNP the residual of a regression on the previous two SNPs was input to the principal component analysis (see EIGENSOFT POPGEN README).

## TreeMix analysis

TreeMix [25] models the genetic drift at genome-wide polymorphisms to infer relationships between populations. It first estimates a dendrogram of the relationships between sampled populations. Next it compares the covariance structure modeled by this dendrogram to the observed covariance between populations. When populations are more closely related than modeled by a bifurcating tree it suggests that there has been admixture in the history of those populations. TreeMix then adds an edge to the phylogeny, now making it a phylogenetic network. The position and direction of these edges are informative; if an edge originates more basally in the phylogenetic network it indicates that this admixture occurred earlier in time or from a more diverged population. TreeMix was used to create a maximum likelihood phylogeny of the 134 breeds. Because TreeMix was slow to add migration events (modeled as “edges”) to the complete data set of 134 breeds, we also analyzed subsets of the data containing considerably fewer breeds. For these subsets, breeds with fewer than 4 samples were removed. To speed up the analysis, we iteratively used the previous graph with $m$-1 migrations as the starting graph and added one migration edge (parameter m set to 1) for a total of $m$ migrations. We rooted the graphs with Bali cattle, used blocks of 1000

SNPs, and used the -se option to calculate standard errors of migration proportions. Migration edges were added until 99.8% of the variance in ancestry between populations was explained by the model. We also ensured that the incorporated migration edges were statistically significant. To further evaluate the consistency of migration edges, we ran TreeMix five separate times with -m set to 17.

### Admixture analysis

ADMIXTURE 1.21 was used to evaluate ancestry proportions for $K$ ancestral populations [26]. We ran ADMIXTURE with cross-validation for values of $K$ from 1 through 20 to examine patterns of ancestry and admixture in our data set.

### $f_3$ and $f_4$ statistics

The $f_3$ and $f_4$ statistics are used to detect correlations in allele frequencies that are not compatible with population evolution following a bifurcating tree; these statistics provide support for admixture in the history of the tested populations [27,28]. The THREEPOP program from TreeMix was used to calculate $f_3$ statistics [28] for all possible triplets from the 134 breeds. The FOURPOP program of TreeMix was used to calculate $f_4$ statistics for subsets of the breeds.

## Results and Discussion

### *Worldwide patterns*

We used principal component analysis (PCA) [24], ancestry graphs implemented in TreeMix [25], and ancestry models implemented in ADMIXTURE [26] to analyze the relationships between 134 breeds of domesticated bovids (Table S1). These breeds arose from three domesticated (sub)species: *Bos javanicus*, *Bos taurus indicus* and *Bos* *taurus taurus* (we use the terms breed and population interchangeably, due to the different definitions of breed worldwide). The principal source of SNP genotype variation was between *Bos t. taurus* and *Bos t. indicus* breeds (Figure 1). This split corresponds to the cattle which originated from the two separate major centers of domestication in the Fertile Crescent and Indus Valley [29]. Although *Bos javanicus* has a more distant common ancestor compared with *Bos t. indicus* and *Bos t. taurus* [3], the uneven sample sizes and ascertainment of SNPs common in *Bos t. taurus* in the design of the BovineSNP50 assay [30] caused the *Bos t. indicus*/*Bos t. taurus* split to be the main source of variation in these data. The second principal component split African taurine cattle from Eurasian taurine, indicine, and Bali cattle.

Early farmers were able to expand their habitat range because of the availability of a reliable supply of food and likely displaced indigenous hunter-gatherer populations by introducing new diseases [31]. The genomes of modern cattle reflect the history of animal movements by migratory farmers out of the ancient centers of cattle domestication. We first ran TreeMix with all 134 populations to identify patterns of divergence (Figure 2). We next ran TreeMix with 74 representative populations (Figure 3, residuals presented in Figure S1) and began to add migration edges to the phylogenetic model (Figure 4, residuals presented in Figure S2). The proportion of the variance in relatedness between populations explained by the model began to asymptote at 0.998 (a value also obtained by simulations [25]) when 17 migration edges were fit (Figure S3). The consistency of these migration edges was evaluated using 5 runs of TreeMix with different random seeds using 17 migration edges (Figure S4). In addition to the migratory routes previously described from the Fertile Crescent to Europe [3], we now find strong evidence of exportations from the Indian subcontinent to China and southeast Asia, India to Africa, Africa to the Iberian peninsula and Mediterranean Europe, India to the Americas, and Europe to the Americas (Figures 4 and 5, discussed in detail in the following subsections). Subsequent to these initial exportations, there have been countless exportations and importations of cattle worldwide. When domesticated cattle were present and new germplasm was imported, the introduced cattle were frequently crossed with the local cattle resulting in an admixed population. Admixed populations were most readily identified when *Bos t. indicus* and *Bos t. taurus* animals were hybridized, which occurred in China, Africa, and the Americas (crosses in Figure 1).

In the late 18th and 19th centuries, European cattlemen began forming closed herds which they developed into breeds [32]. Because breeds are typically reproductively isolated with little or no interbreeding, we found that the cross-validation error estimates continued to decrease as we increased the number of ancestral populations $K$ modeled in the admixture analysis (Table S2). This reflects the large differences in allele frequencies that exist between breeds resulting from separate domestication events, geographic dispersal and isolation, breed formation, and the use of artificial insemination. The method of Evanno et al. [33], which evaluates the second order rate of change of the likelihood function with respect to $K$ ($\Delta K$), identified $K = 2$ as the optimum level of $K$ (Figure S5). This method was overwhelmed by the early divergence between indicine and taurine cattle, and was not sensitive to the hierarchical relationships of populations and breeds [34]. As we increased the value of *K*, we recapitulated the increasingly fine structure represented in the branches of the phylogeny (Figures 6, S6-S10).

### *Modern Anatolian cattle are not representatives of early domesticated cattle*

Anatolian breeds (AB, EAR, TG, ASY, and SAR) are admixed between blue European-like, grey African-like, and green indicine-like cattle (Figures 5 and 6), and we infer that they do not represent the taurine populations originally domesticated in this region due to a history of admixture. Zavot (ZVT), a crossbred breed [32], has a different history with a large portion of ancestry similar to Holsteins (Figures 2 and S8-S10). The placement of Anatolian breeds along principal components 1 and 2 in Figure 1 [30], the ancestry estimates in Figure 6, their extremely short branch lengths in Figures 2-4, and significant $f_3$ statistics confirm that modern Anatolian breeds are admixed. For example, the Anatolian Southern Yellow (ASY) has 3,003 significant $f_3$ tests, the most extreme of which has Vosgienne (VOS, a taurine breed) and Achai (ACH, an indicine breed) as sister groups with a Z-score of -43.69. Our results support previous work using microsatellite loci [35] which inferred Anatolian cattle to possess indicine introgression. We further demonstrate that Anatolian breeds are admixed between European and African. We calculated $f_4$ statistics with East Anatolian Red, Anatolian Southern Yellow, and Anatolian Black as sister, and N'Dama, Somba, Lagune, Baole, Simmental, Holstein, Hereford, and Shorthorn as the opposing sister group. From Figure 2, we would expect these relationships to be tree-like. But 45 of the possible 84 $f_4$ tests indicated significant levels of admixture. The most significant was $f_4$(East Anatolian Red, Anatolian Southern Yellow; Somba, Shorthorn) = -0.0026 ± 0.0003 (Z-score = -8.10).

***Divergence within the taurine lineage***

If African and Asian taurines were both exported from the Fertile Crescent in similar numbers at about the same time, we would expect them to be approximately equally diverged from European taurines. However, African taurines were consistently revealed to be more diverged from European and Asian taurines (Figures 1, 2, 3, and 5, Anatolian breeds are not considered in this comparison because of their admixed history). Two factors appear to influence this divergence. First, European cattle were exported into Asia and admixed with Asian taurines. In the admixture models in which $K$ = 15 or 20 (Figures S9 and S10), there was evidence of European taurine admixture in the Mongolian (MG), Hanwoo (HANW), and Wagyu (WAGY) breeds. We ran TreeMix with 14 representative populations and estimated Wagyu to have 0.188 ± 0.069 (p-value = 0.003) of their genome originating from northwestern European ancestry (Figure 7). We also see some runs of TreeMix placing a migration edge from Chianina cattle to Asian taurines (Figure S4). We ran $f_4$ tests with Mongolian, Hanwoo, Wagyu, Tharparkar (THA), or Kankraj (KAN) as sister populations, and Piedmontese (PIED), Simmental (SIM), Brown Swiss (BSW), Braunvieh (BRVH), Devon (DEV), Angus (AN), Shorthorn (SH), or Holstein (HO) as the opposing pair of sister groups. From previous research [3] and Figures 2 and 3, these relationships should be tree-like if there were no admixture. For 53 of the possible 280 tests, the Z-score was more extreme than ±2.575829. The most extreme test statistics were $f_4$(Wagyu, Mongolian; Simmental, Shorthorn) = -0.003 (Z-score = -5.21, other rearrangements of these groups had Z-scores of 7.32 and 16.55) and $f_4$(Hanwoo, Wagyu; Piedmontese, Shorthorn) = 0.002 (Z-score = 4.90, other rearrangements of these groups had Z-scores of 21.79 and 27.77). When $K$ = 20, Hanwoo appear to have a Mediterranean influence, whereas Wagyu have a northwestern European, including British, influence (Figure S10). We conclude that there were two waves of European introgression into Far East Asian cattle, first with Mediterranean cattle (which carried African taurine and indicine alleles) brought along the Silk Road and later from 1868 to 1918 when Japanese cattle were crossed with British and Northwest European cattle [32].

The second factor that we believe underlies the divergence of African taurine is a high level of wild African auroch [36,37] introgression. Principal component (Figure 1), phylogenetic trees (Figures 2 and 3), and admixture (Figure 6) analyses all reveal the African taurines as being the most diverged of the taurine populations. Because of this divergence, it has been hypothesized that there was a third domestication of cattle in Africa [38–42]. If there was a third domestication, African taurine would be sister to the European and Asian clade. When no migration events were fit in the TreeMix analyses, African cattle were the most diverged of the taurine populations (Figures 2 and 3), but when admixture was modeled to include 17 migrations, all African cattle, except for East African Shorthorn Zebu and Zebu from Madagascar which have high indicine ancestry, were sister to European cattle and were less diverged than Asian or Anatolian cattle (Figure 4), thus ruling out a separate domestication. Our phylogenetic network (Figure 4) shows that there was not a third domestication process, rather there was a single origin of domesticated taurine (Asian, African, and European all share a recent common ancestor denoted by an asterisk in Figure 4, with Asian cattle sister to the rest of the taurine lineage), followed by admixture with an ancestral population in Africa (migration edge *a* in Figure 4, which is consistent across 6 separate TreeMix runs, Figure S4). This ancestral population (origin of migration edge *a* in Figure 4) was approximately halfway between the common ancestor of indicine and the common ancestor of taurine. We conclude that African taurines received as much as 26% (estimated as 0.263 in the network, p-value < 2.2e-308) of their ancestry from admixture with wild African auroch. Although three other migration edges originate from the branch between indicine and taurine (such as edge *b*), all of the receiving populations show indicine ancestry in the ADMIXTURE models. But African auroch are extinct and samples were not available for the ADMIXTURE model, thus the admixed auroch ancestry of African taurines cannot specifically be discovered by this model [34,43] and African taurine, especially Lagune, are depicted as having a single ancestry without indicine influence (Figures 5 and 6, see $f_3$ and $f_4$ statistics reported later). Unlike ADMIXTURE, TreeMix can model admixture from an unsampled population by placing a migration edge more basal along a branch of the phylogeny, in this case African auroch.

Others have observed distinct patterns of linkage disequilibrium in African taurines, resulting in larger estimates of ancestral effective population size than for either *Bos t. taurus* or *Bos t. indicus* breeds [2] consistent with greater levels of admixture from wild aurochs. Just as Near Eastern domesticated pig mitochondrial lineages were replaced by mitochondria from indigenous wild populations [44], we infer that the divergent

African mitochondrial sequences [38] previously observed result from admixture with wild African auroch. Similar patterns of admixture from wild forebears have been observed in other species [44], such as pig [45–47], chicken [48], and corn [14], and this conclusion represents the most parsimonious explanation of our results. We hypothesize that the auroch introgression in Africa may have been driven by trypanosomiasis resistance in African auroch which may be the source of resistance in modern African taurine populations [49]. Admixture with distant relatives has had an important impact on the immune system of other species, such as human [50] and possibly chicken [51]. More sophisticated demographic models and unbiased whole-genome sequence data will be needed to further test these hypotheses.

***Indicine admixture in Africa***

African cattle also demonstrate a geographical gradient of indicine ancestry [52]. Taurine cattle in western Africa possess from 0% to 19.9% indicine ancestry (Figures 5 and 6, LAG, ND1, ND2, NDAM, BAO, OUL, SOM), with an average of 3.3%. Moving from west to east and from south to central Africa, the percent of indicine ancestry increases from 22.7% to 74.1% (Figures 5 and 6, ZFU, ZBO, ZMA, BORG, TULI, BOR, SHK, ZEB, ANKW, LAMB, an AFR), with an average of 56.9%. As we increased values of $K$ to 10, 15, and 20 (Figures S8-S10), we revealed two clusters of indicine ancestry possibly resulting from the previously suggested two waves of indicine importation into Africa, the first occurring in the second millennium BC and the second during and after the Islamic conquests [32,40,53]. The presence of two separate clades of African cattle in Figure 4 also supports the idea of two waves of indicine introgression.

***Admixture in Asia***

Asian cattle breeds were derived from cattle domesticated in the Indian subcontinent or imported from the Fertile Crescent and Europe. Cattle in the north and northeast are primarily of *Bos t. taurus* ancestry (Figures 5 and 6; HANW, WAGY, and MG), but Hanwoo and Mongolian also have *Bos t. indicus* ancestry (Figures 5, 6, S9, and S10). Cattle in Pakistan, India, southern China and Indonesia are predominantly *Bos t. indicus* (Figures 5 and 6; ONG, MAD, BRE, HN, ACE, PES, ACH, HAR, BAG, GUZ, SAHW, GBI, CHO, GIR, KAN, THA, RSIN, HIS, LOH, ROJ, DHA, and DAJ). Cattle located between these two geographical regions are *Bos t. taurus* × *Bos t. indicus* hybrids (Figures 1, 4, 5, and 6; QC and LX). Our results suggest an additional source for increased indicine diversity—admixture with domesticated cattle from other species. In addition to cattle domesticated from aurochs (*Bos primigenius*), bovids were also domesticated from water buffalo (*Bubalus bubalis*), yak (*Bos grunniens*), gaur (*Bos gaurus*), and banteng (*Bos javanicus*), represented in our sample by the Bali breed [32,54]. We find that the Indonesian Brebes (BRE) and Madura (MAD) breeds have significant *Bos javanicus* (BALI) ancestry demonstrated by the short branch lengths in Figures 2-4, shared ancestry with Bali in ADMIXTURE analyses (light green in Figures S8-S10), and significant $f_3$ statistics (Table S3). The Indonesian Pesisir and Aceh and the Chinese Hainan and Luxi breeds also have Bali ancestry (migration edge *c* in Figure 4, migration edges in Figure S4, and light green in Figures S8 and S9).

***Admixture in Europe***

Cattle were imported into Europe from the southeast to the northwest. The descendants of Durham Shorthorns (the ancestral Shorthorn breed [32]) were the most distinct group of European cattle as they clustered at the extremes of principal component 2 (lower left hand corner of Figure 1), and they formed a distinct cluster in the ADMIXTURE analyses whenever *K* was greater than 4 (Figures S6-S10). As shown in Figures S6 through S10, $f_3$ statistics in Table S4, and from their breed histories [32], many breeds share ancestry with Shorthorn cattle, including Milking Shorthorn, Beef Shorthorn, Lincoln Red, Maine-Anjou, Belgian Blue, Santa Gertrudis, and Beefmaster.

From the previous placement of the American Criollo breeds including Romosinuano, Texas Longhorn, and Corriente, it has been posited that Iberian cattle became admixed as a result of an introgression of cattle from Africa into the local European cattle [3,55,56]. Our genotyping of individuals from 11 Spanish breeds supported, but clarified, this hypothesis. On average, Spanish cattle had 19.3% of African ancestry when $K = 3$, with a minimum of 8.8% and a maximum of 23.4%, which supports previous analyses of mitochondrial DNA [57,58]. Migration edge *d* in the phylogenetic network (Figure 4, and consistently seen in Figure S4) estimates that Iberian cattle, Texas Longhorn, and Romosinuano derive 7.5% of their ancestry from African taurine introgression, similar to the ancestry estimates from the models with larger *K* values (Figures S8-S10). The Oulmès Zaer (OUL) breed from Morocco also shows that cattle were transported from Iberia and France to Africa (tan and red in Figure S10, and short branch length in Figure 4). However, the 11 Spanish breeds had no more indicine ancestry than all other European taurine breeds (essentially none for the majority of breeds, see Figures 5 and 6). Maraichine (MAR), Gascon (GAS), Limousin (LIM), and other breeds from France, and Piedmontese cattle (PIED) from northwest Italy have a similar ancestry. These data indicate that the reason that the American Criollo breeds were found to be sister to European cattle in our previous work [3] was because of their higher proportion of indicine ancestry. The 5 sampled American Criollo breeds had, on average, 14.7% African ancestry (minimum of 6.2% and maximum of 20.4%) and 8.0% indicine ancestry (minimum of 0.6% and maximum of 20.3%).

Other Italian breeds (MCHI, CHIA, and RMG) share ancestry with both African taurine and indicine cattle (Figures 6, S6-S8). This introgression may have come from Anatolian or East African cattle that carried both African taurine and indicine ancestry, which is modeled as migration edge *b* in Figure 4. The placement of Italian breeds is not consistent across independent TreeMix runs (Figure S4), likely due to their complicated history of admixture.

We also used *f*-statistics to explore the evidence for African taurine introgression into Spain and Italy. We did not see any significant $f_3$ statistics, but this test may be underpowered because of the low-level of introgression. With Italian and Spanish breeds as a sister group and African breeds, including Oulmès Zaer, as the other sister group, we see 321 significant tests out of 1,911 possible tests. Of these 321 significant tests, 218 contained Oulmès Zaer. We also calculated $f_4$ statistics with the Spanish breeds as sister and the African taurine breeds as sister (excluding Oulmès Zaer). With this setup, out of the possible 675 tests we saw only 1 significant test, $f_4$(Berrenda en

Negro, Pirenaica;Lagune, N'Dama (ND2)) = 0.0007, Z-score = 3.064. With Italian cattle as sister and African taurine as sister (excluding Oulmès Zaer), we saw 17 significant tests out of the 90 possible. Patterson et al. [28] defined the $f_4$-ratio as $f_4$(A, O; X, C)/$f_4$(A, O; B, C), where A and B are a sister group, C is sister to (A,B), X is a mixture of B and C, and O is the outgroup. This ratio estimates the ancestry from B, denoted as $\alpha$ , and the ancestry from C, as 1 - $\alpha$ . We calculated this ratio using Shorthorn as A, Montbeliard as B, Lagune as C, Morucha as X, and Hariana as O. We choose Shorthorn, Montbeliard, Lagune, and Hariana as they appeared the least admixed in the ADMIXTURE analyses. We choose Morucha because it appears as solid red with African ancestry in Figure S10. This statistic estimated that Morucha is 91.23% European ($\alpha$ = 0.0180993/0.0198386) and 8.77% African, which is similar to the proportion estimated by TreeMix. The multiple $f_4$ statistics with Italian breeds as sister and African breeds as the opposing sister support African admixture into Italy. The $f_4$-ratio test with Morucha also supports our conclusion of African admixture into Spain.

***Preservation of pure taurine in Africa and lack of widespread indicine ancestry in Europe***

It has recently been concluded that indicine ancestry is a common feature of European cattle genomes [59]. However, our data refute this conclusion. McTavish et al. relied on the Evanno test to arrive at an optimal number of ancestral populations of $K$ = 2, which masks the fact that there are cattle breeds in Africa with 100% African taurine ancestry (Figure 6). Although our $K$ = 2 ADMIXTURE results suggested that most African breeds had at least 20% indicine ancestry (Figure S5), when we increased $K$ to 3, Lagune (LAG) revealed no indicine ancestry, and Baoule (BAO) and N'Dama (NDAM) possess very little indicine ancestry. If the *K* = 2 model was correct, we would expect to see numerous significant $f_3$ and $f_4$ tests with Eurasian taurine and indicine as sister groups. Whereas, if the *K* = 3 model more accurately reflected the heritage of European and African taurines, we would not observe any significant $f_3$ or $f_4$ tests showing admixture of taurine and indicine in the ancestry of African taurine. For the Lagune, Baoule and N'Dama (NDAM and ND2) breeds we found no significant $f_3$ statistics. Among the 225 $f_4$ statistics calculated with NDAM, LAG, BAO, ND2, SH, and MONT as sisters and BALI, GIR, HAR, SAHW, PES, and ACE as the opposing sister group, only 36 were significantly different from 0 (Table 1). When ND2 was excluded from the results, only 4 tests were significant (Table 1), and we have no evidence that the Lagune breed harbors indicine alleles. Thus, we conclude that contrary to the assumptions and conclusions of [59] cattle with pure taurine ancestry do exist in Africa. Further, we conclude that indicine ancestry in European taurine cattle is extremely rare, and that some breeds, especially those prevalent near the Mediterranean, possess African taurine introgression—but with the exception of the Charolais, Marchigiana, Chianina and Romagnola breeds—not African hybrid or African indicine introgression. We concur that Texas Longhorn and other American Criollo breeds possess indicine ancestry, but infer that this introgression occurred after the arrival of Spanish cattle in the New World and likely originated from Brahman cattle (migration edges *e* and *f* in Figure 4). In TreeMix replicates, Texas Longhorn and Romosinuano are either sister to admixed Anatolian breeds or they receive a migration edge that originates near Brahman (Figure S4). To reiterate, Iberian cattle do not have indicine ancestry, American Criollo breeds originated from exportations from Iberia, Brahman cattle were developed in the United States in the 1880's [32], American Criollo breeds carry indicine ancestry, and the introgression likely occurred from Brahman cattle.

Domestication, exportation, admixture, and breed formation have had tremendous impacts on the variation present within and between cattle breeds. In Asia, Africa, North and South America, cattle breeders have crossbred *Bos t. taurus* and *Bos t. indicus* cattle to produce hybrids which were well suited to the environment and endemic production systems. In this study, we clarify the relationships between breeds of cattle worldwide, and present the most accurate cattle "Tree of Life" to date in Figure 4. We elucidate the complicated history of Asian cattle involving the domestication and subsequent admixture of several bovid species. We provide evidence for admixture between domesticated Fertile Crescent taurine and wild African auroch in Africa to form the extant African taurine breeds. We also observe African taurine content within the genomes of European Mediterranean taurine breeds. The absence of indicine content within the majority of European taurine breeds, but the presence of indicine within three Italian breeds is consistent with two separate introductions, one from the Middle East potentially by the Romans which captured African taurines in which indicine introgression had already occurred and the second from western Africa into Spain which included African taurines with no indicine introgression. It was this second group of cattle which likely radiated from Spain into Southern France and the Alps. The prevalence of admixture further convolutes the cryptic history of cattle domestication.

## Acknowledgments

We gratefully acknowledge the provision of samples and genotypes from breed associations, cattle breeders, semen distributors and the Bovine HapMap Project. This project was supported by National Research Initiative grants number 2008-35205-04687 and 2008-35205-18864 from the USDA Cooperative State Research, Education and Extension Service and National Research Initiative grants number 2009-65205-05635, 2011-68004-30214, 2011-68004-30367 and 2013-68004-20364 from the USDA National Institute of Food and Agriculture. We thank Daniel G. Bradley for insightful comments on the analysis and manuscript and Joseph Pickrell and an anonymous reviewer for constructive peer-reviews.

## Figure Legends

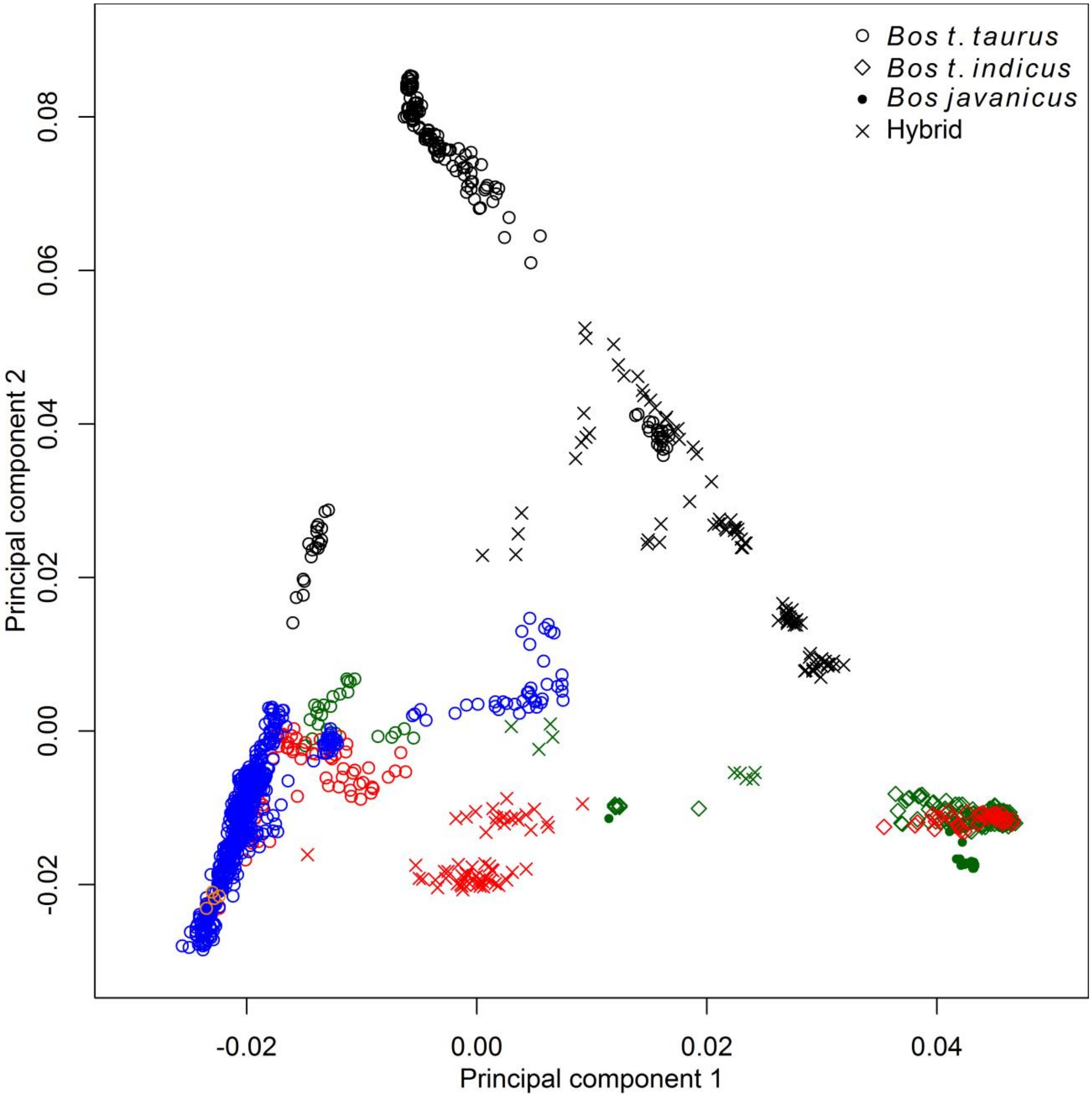

**Figure 1. Principal component analysis of 1,543 animals genotyped with 43,043 SNPs.** Points were colored according to geographic origin of breed; black: Africa, green: Asia, red: North and South America, orange: Australia, and blue: Europe.

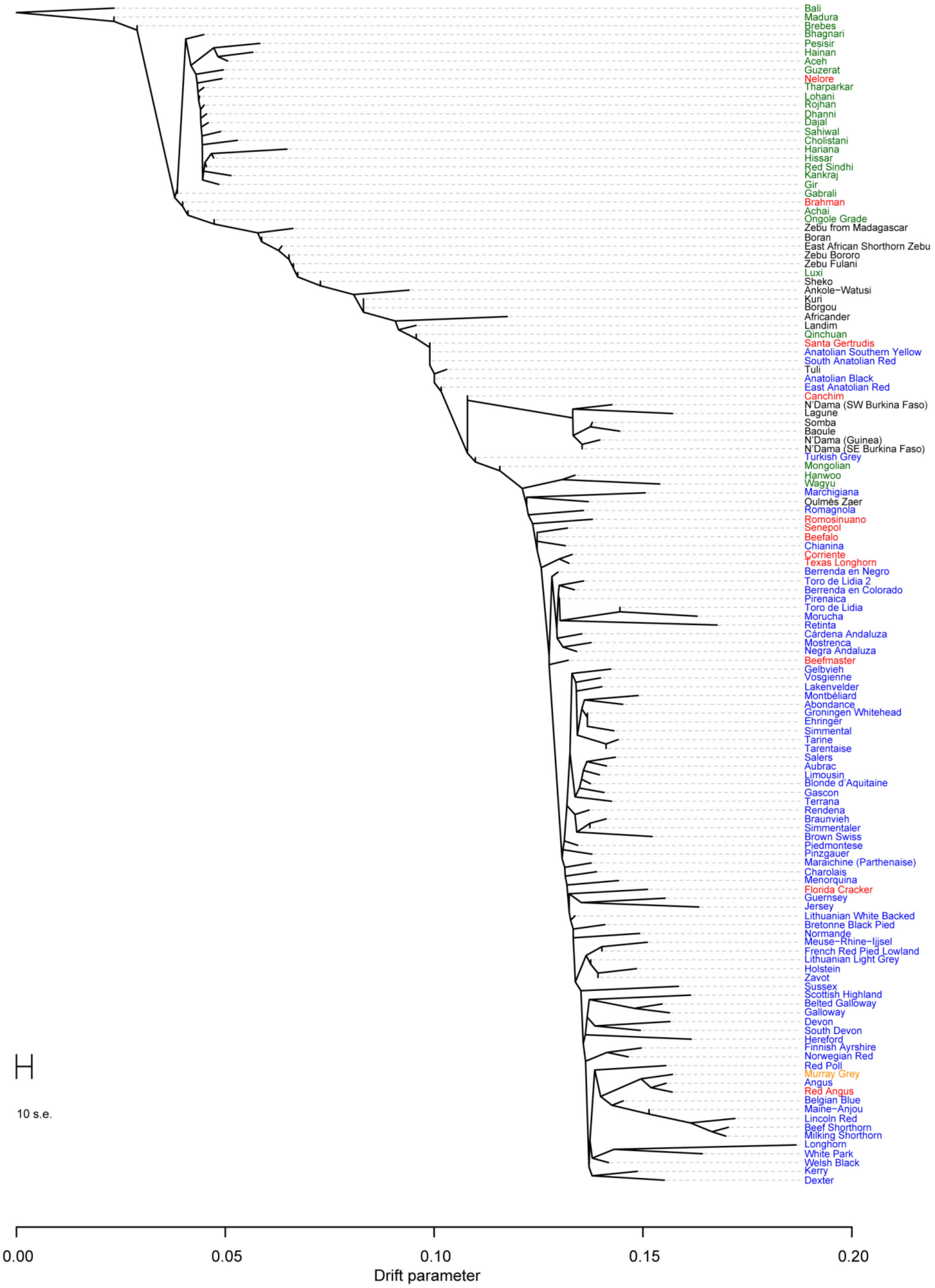

**Figure 2. Phylogram of the inferred relationships between 134 cattle breeds.** Breeds were colored according to their geographic origin; black: Africa, green: Asia, red: North and South America, orange: Australia, and blue: Europe. Scale bar shows 10 times the average standard error of the estimated entries in the sample covariance matrix (See [25]).

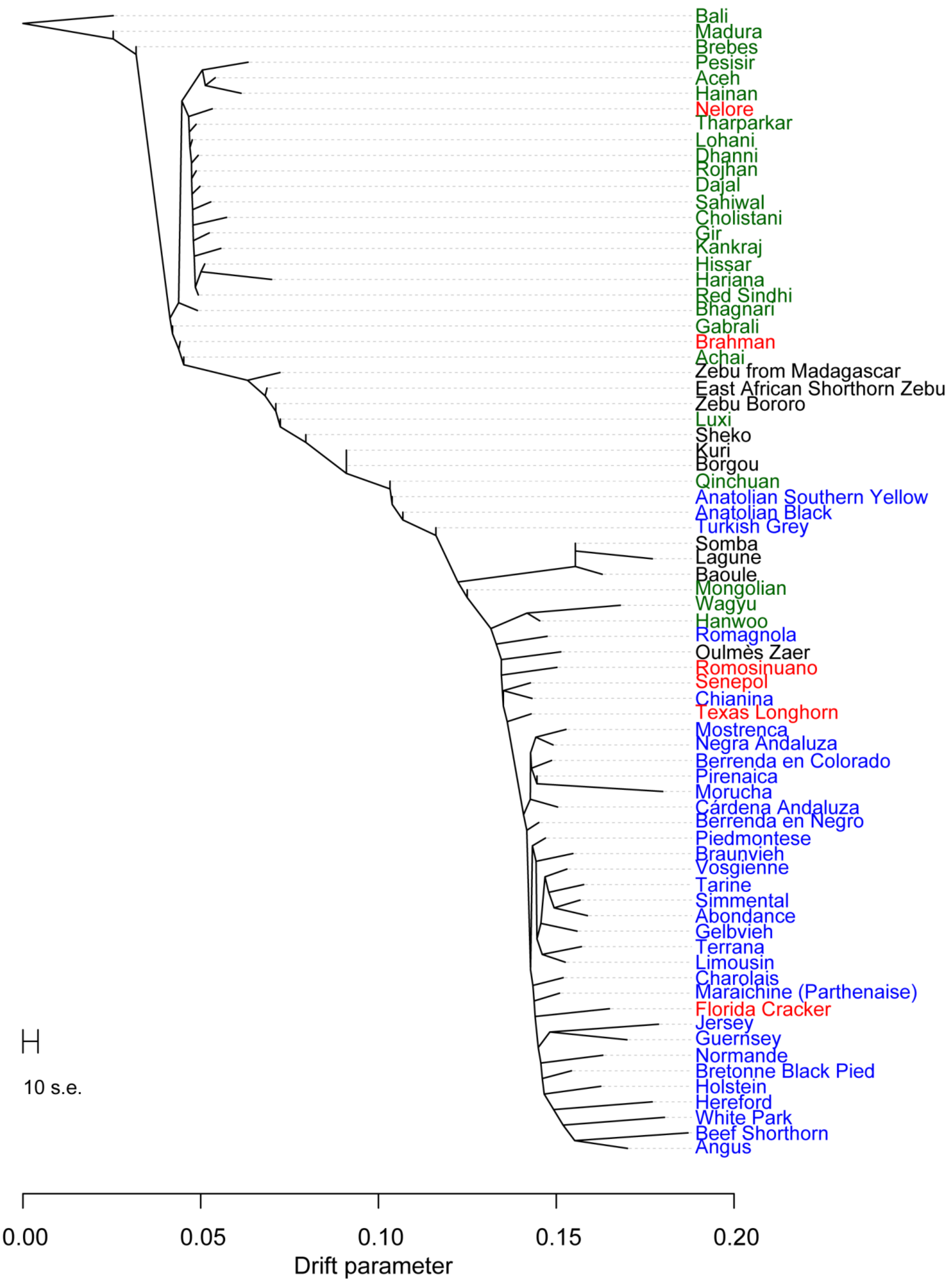

**Figure 3. Phylogram of the inferred relationships between 74 cattle breeds.** Breeds were colored according to their geographic origin; black: Africa, green: Asia, red: North

and South America, orange: Australia, and blue: Europe. Scale bar shows 10 times the average standard error of the estimated entries in the sample covariance matrix.

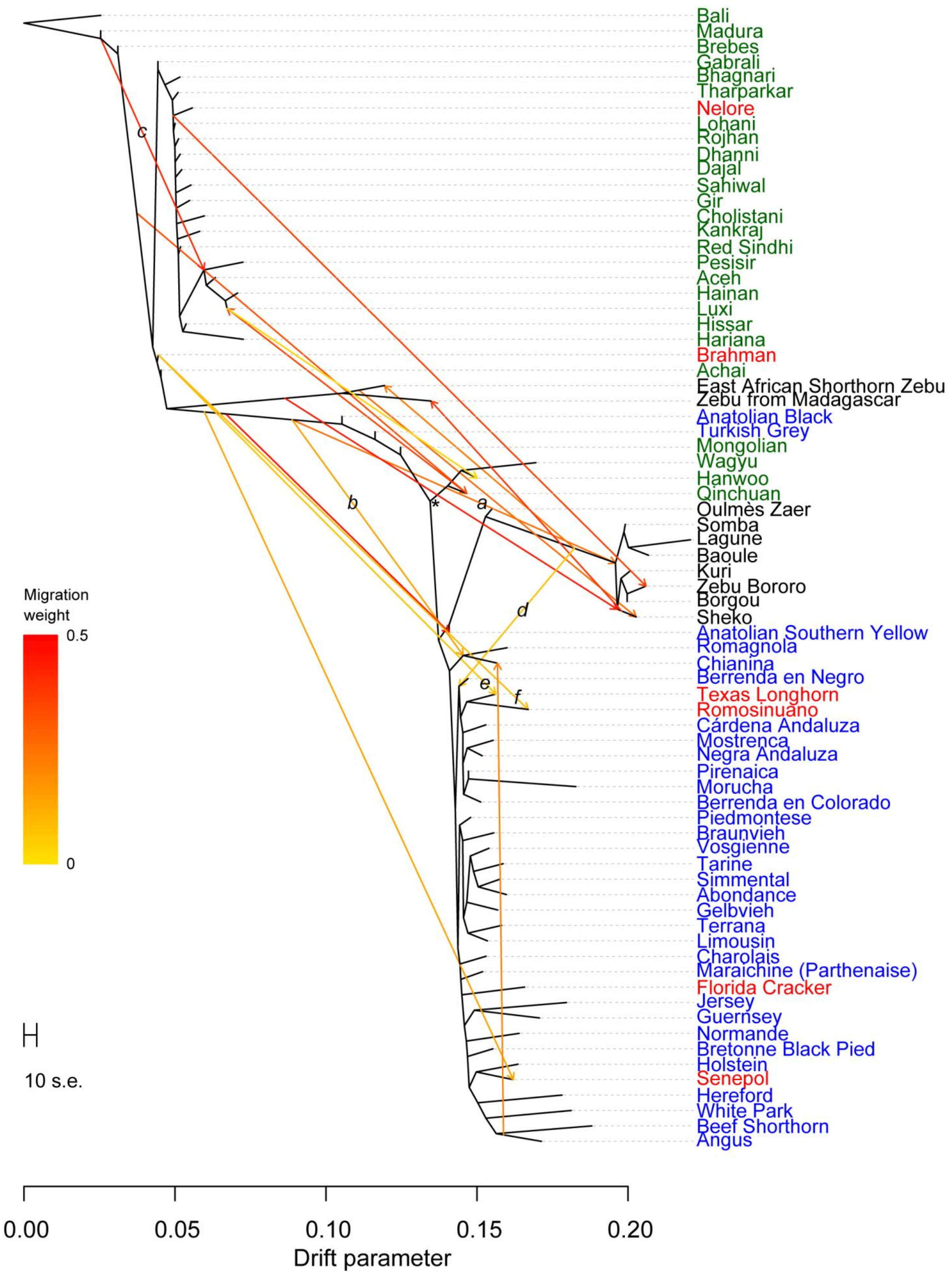

**Figure 4. Phylogenetic network of the inferred relationships between 74 cattle breeds.** Breeds were colored according to their geographic origin; black: Africa, green: Asia, red: North and South America, orange: Australia, and blue: Europe. Scale bar shows 10 times the average standard error of the estimated entries in the sample covariance matrix. Common ancestor of domesticated taurines is indicated by an asterisk. Migration edges were colored according to percent ancestry received from the donor population. Migration edge *a* is hypothesized to be from wild African auroch into domesticates from the Fertile Crescent. Migration edge *b* is hypothesized to be introgression from hybrid African cattle. Migration edge *c* is hypothesized to be introgression from Bali/indicine hybrids into other Indonesian cattle. Migration edge *d* signals introgression of African taurine into Iberia. Migration edges *e* and *f* represent introgression from Brahman into American Criollo.

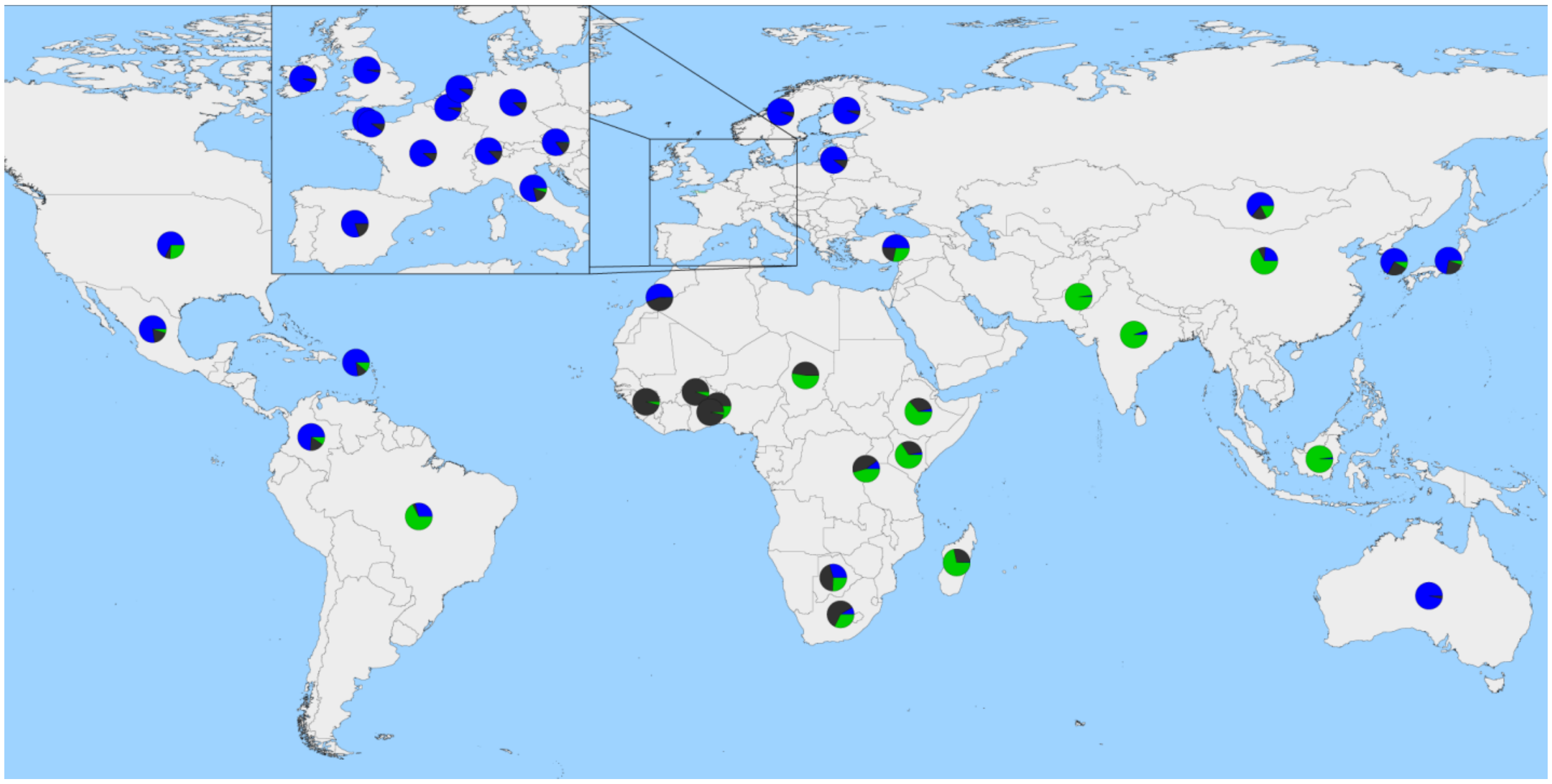

**Figure 5. Worldwide map with country averages of ancestry proportions with 3 ancestral populations (*K* = 3).** Blue represents Eurasian *Bos t. taurus* ancestry, green represents *Bos javanicus* and *Bos t. indicus* ancestry, and dark grey represents African *Bos. t. taurus* ancestry. Please note, averages do not represent the entire populations of each country, as we do not have a geographically random sample.

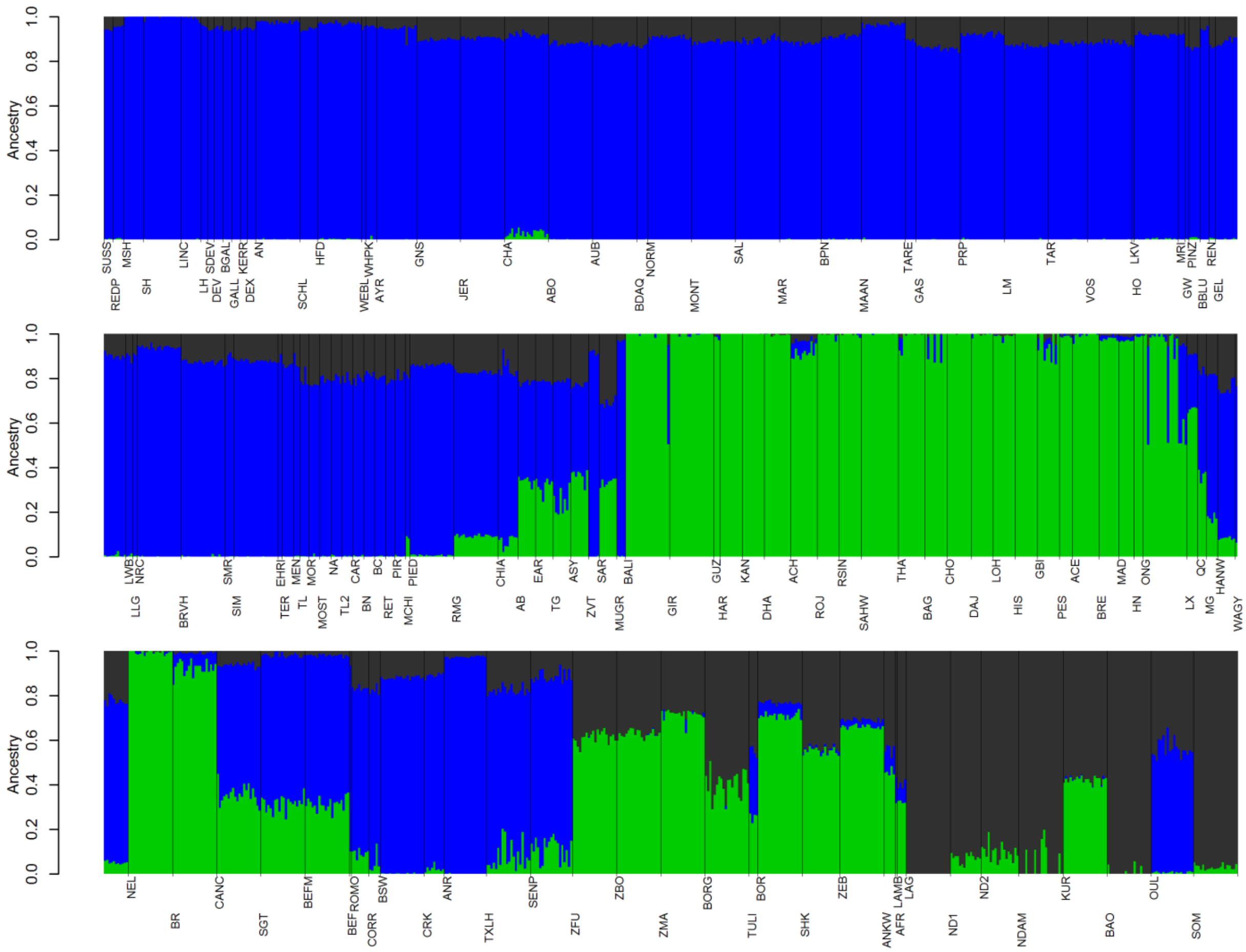

**Figure 6. Ancestry models with 3 ancestral populations (*K* = 3).** Blue represents Eurasian *Bos t. taurus* ancestry, green represents *Bos javanicus* and *Bos t. indicus* ancestry, and dark grey represents African *Bos. t. taurus* ancestry. See Supplementary Figures S5-S10 for other values of *K*.

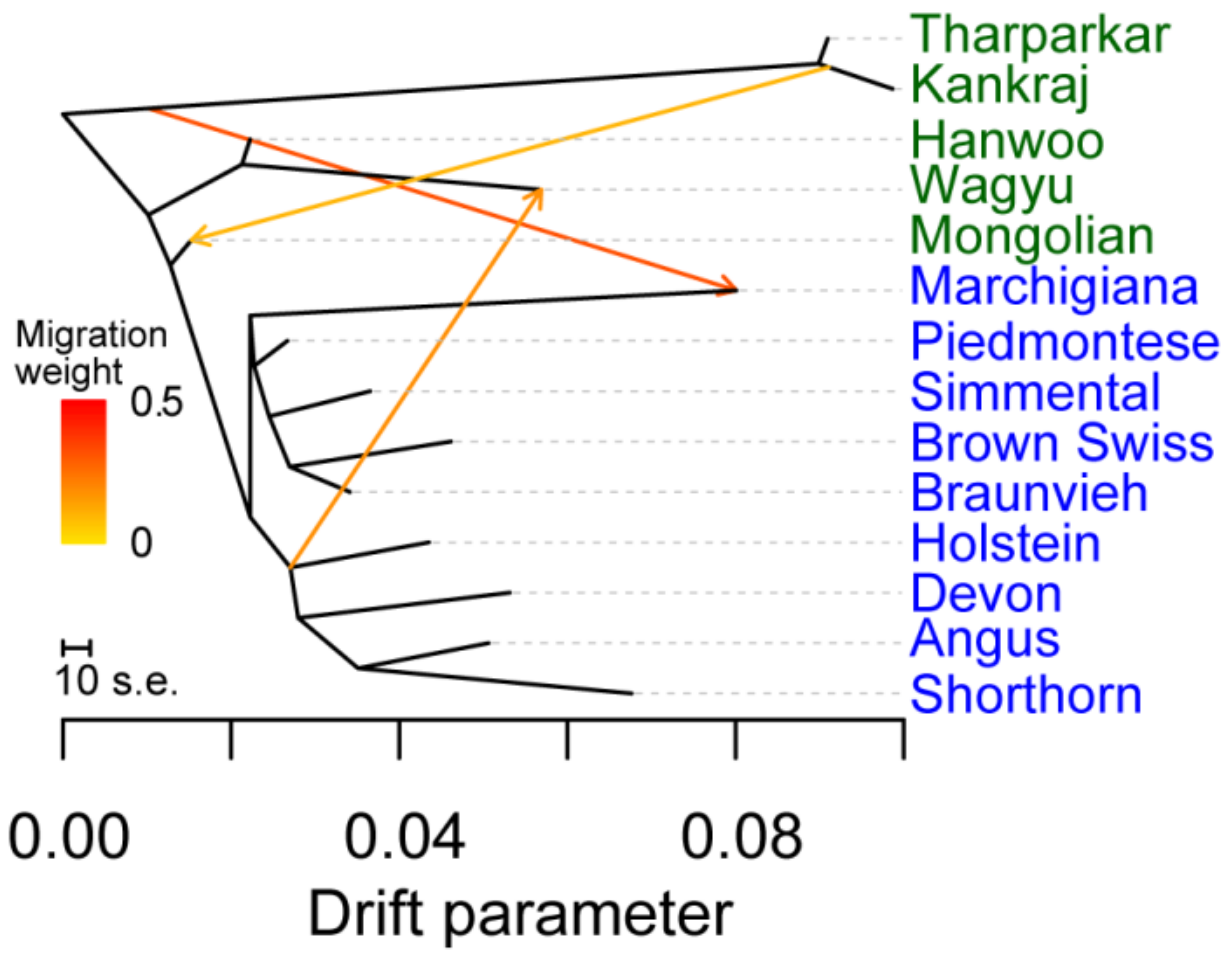

**Figure 7. Phylogenetic network of the inferred relationships between 14 cattle breeds.** Breeds were colored according to their geographic origin; green: Asia, and blue: Europe. Scale bar shows 10 times the average standard error of the estimated entries in the sample covariance matrix. Migration edges were colored according to percent ancestry received from the donor population. Migration edges show indicine introgression into Mongolian cattle, African taurine and indicine ancestry in Marchigiana, and a northern European influence on Wagyu.

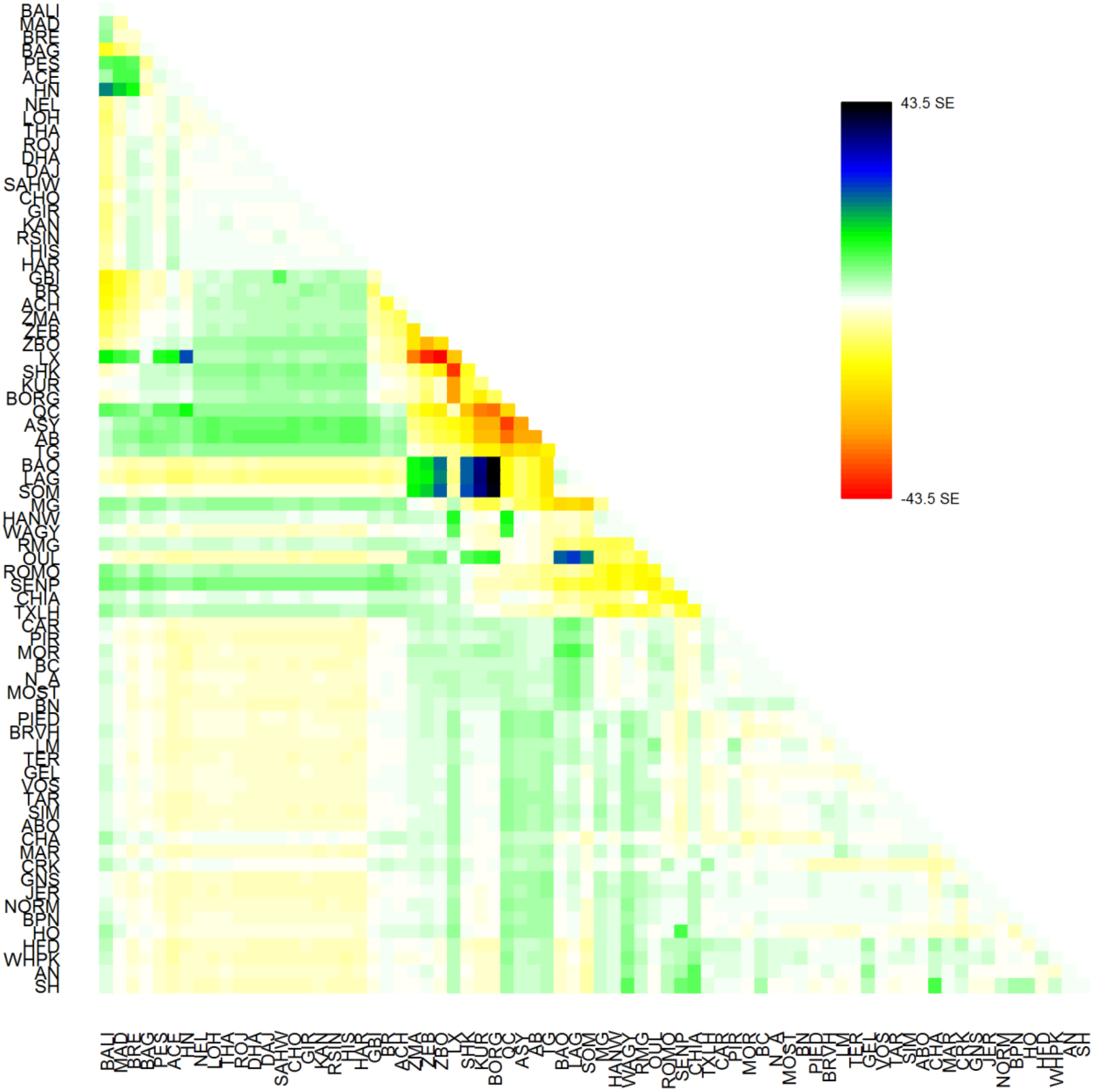

**Figure S1. Plot of residuals from the phylogeny model depicted in Figure 3 when no migration edges were fit.**

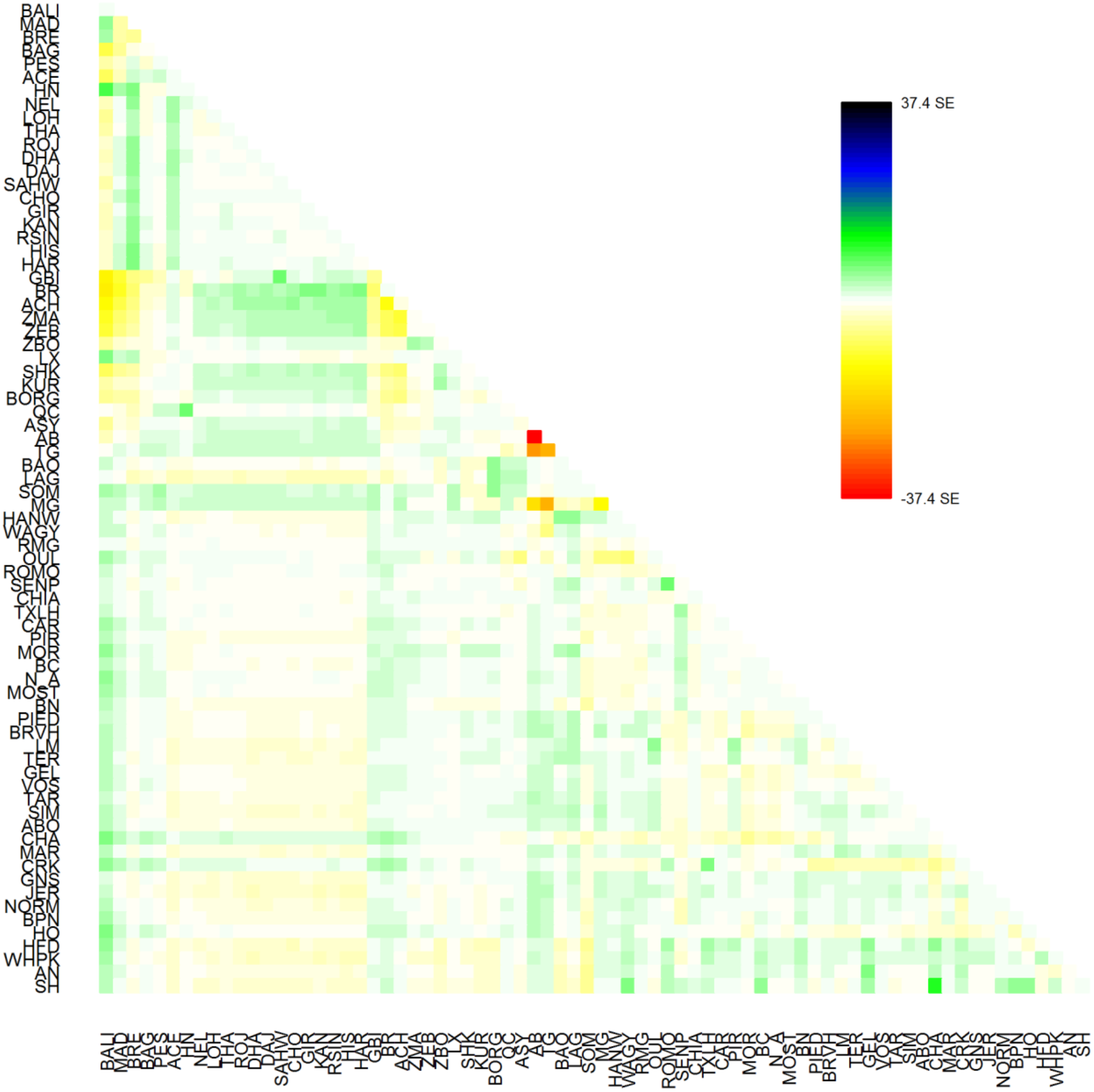

**Figure S2. Plot of residuals from the phylogenetic network model depicted in Figure 4 when 17 migration edges were fit.**

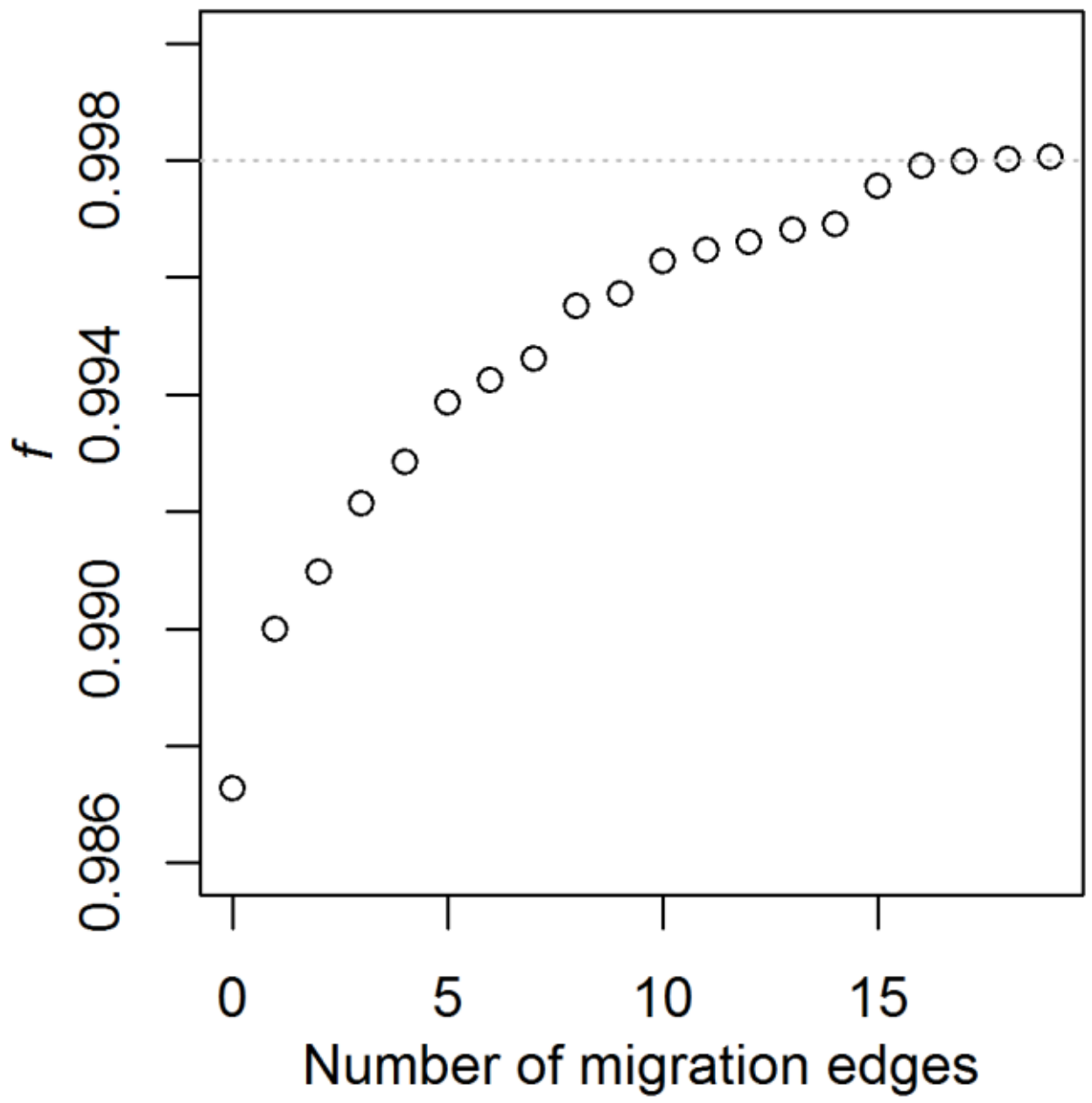

**Figure S3. The fraction of variance in relatedness between populations accounted for by phylogenetic models with 0 through 19 migrations.** The fraction of variance in the sample covariance matrix ($\hat{W}$) accounted for by the model covariance matrix ($W$). Pickrell and Pritchard [25] showed that the fraction began to asymptote at 0.998 when the models accurately depicted relationships between simulated populations. We also observed this asymptote near 0.998 in our empirical analysis, leading us to conclude that the relationships between the 74 cattle breeds were accurately described by a phylogenetic network with 17 migration edges.

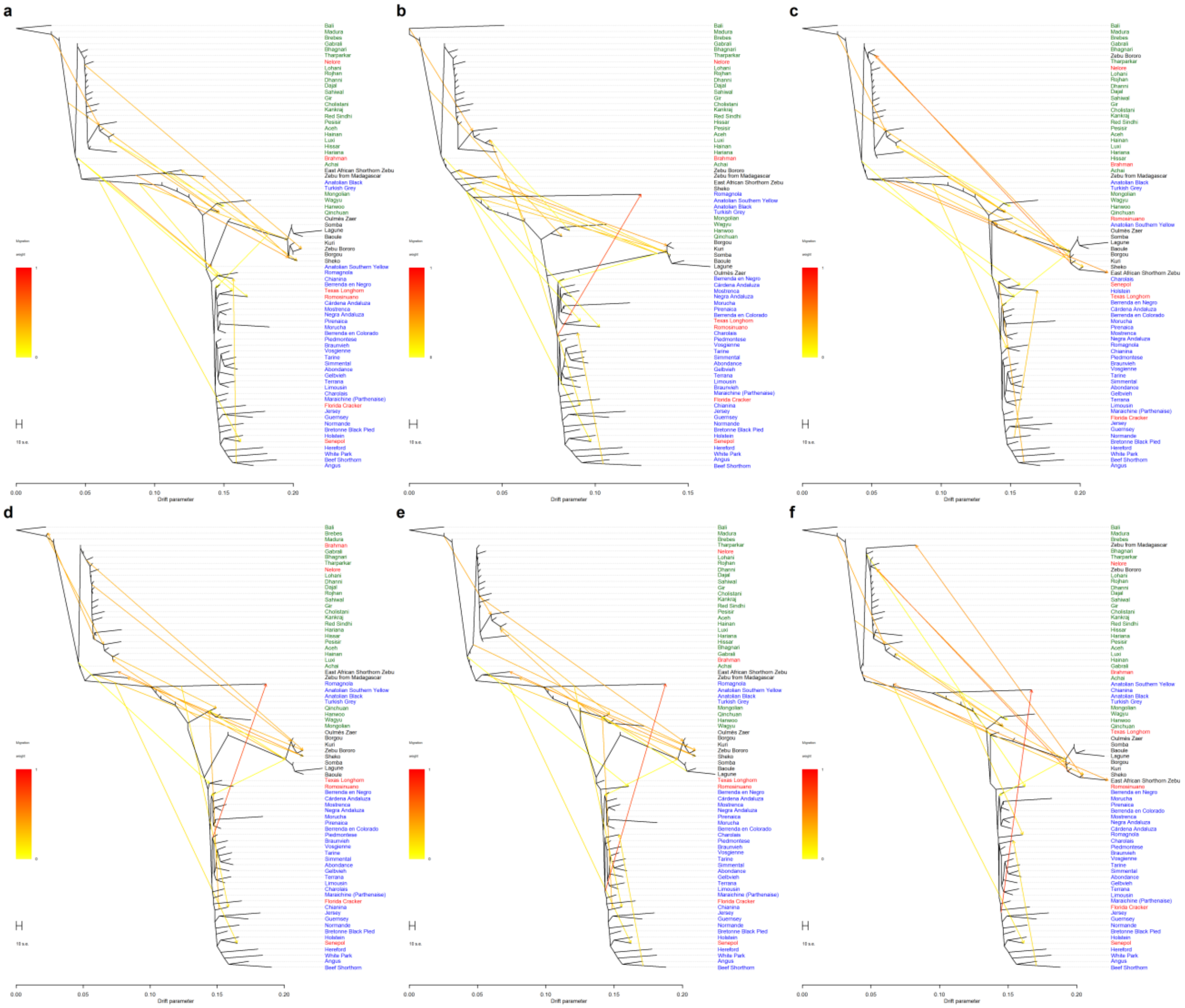

**Figure S4. Phylogenetic network with 17 edges (Figure 4) plus 5 replicates.** Replicates were run with different random seeds to visually evaluate consistency of migration edges. Network a is the same as Figure 4; networks b through f are replicates. Breeds were colored according to their geographic origin; black: Africa, green: Asia, red: North and South America, orange: Australia, and blue: Europe. Scale bar shows 10 times the average standard error of the estimated entries in the sample covariance matrix. Migration edges were colored according to percent ancestry received from the donor population.

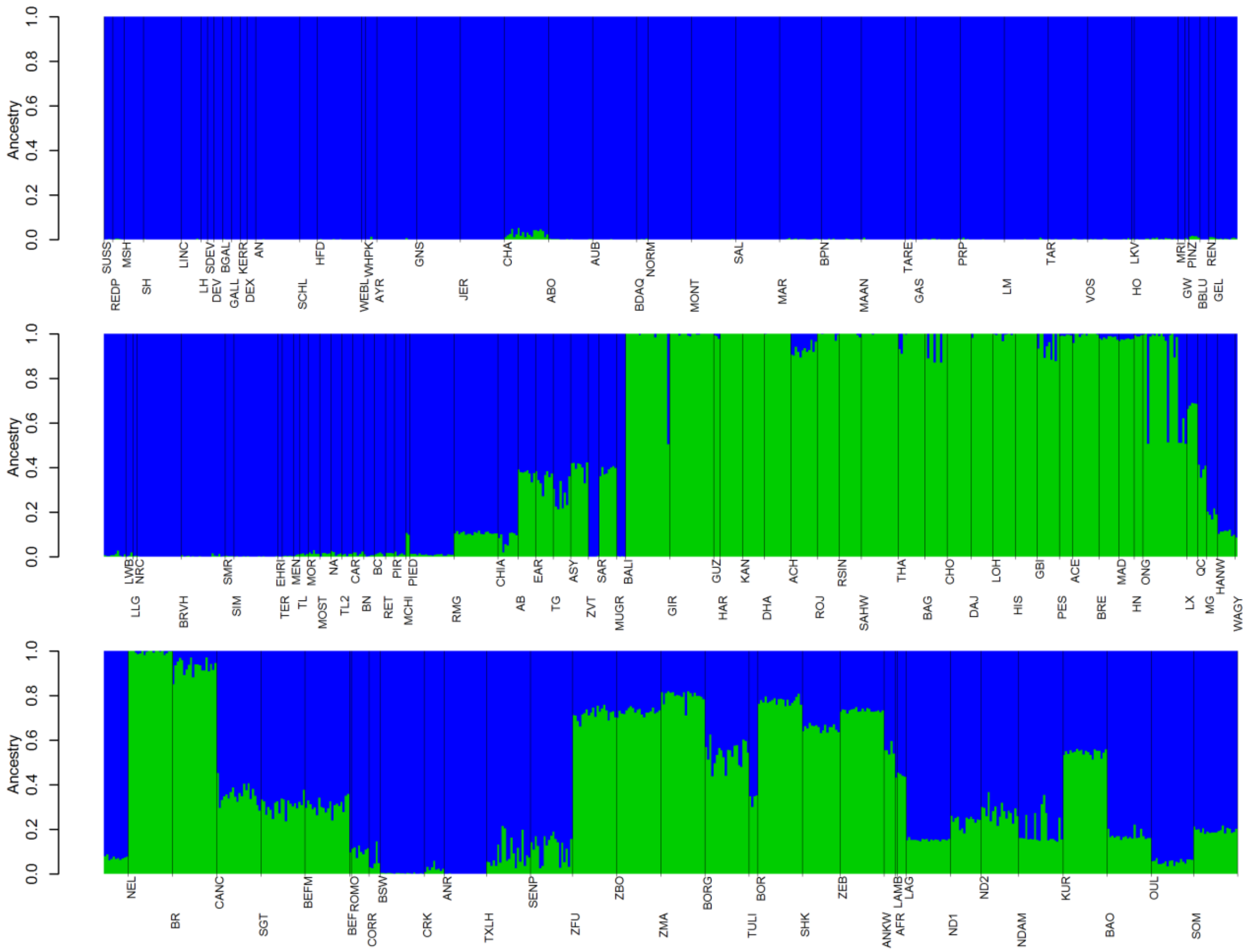

**Figure S5. Ancestry models with 2 ancestral populations (*K* = 2).** Blue represents *Bos t. taurus* ancestry, and green represents *Bos javanicus* and *Bos t. indicus* ancestry.

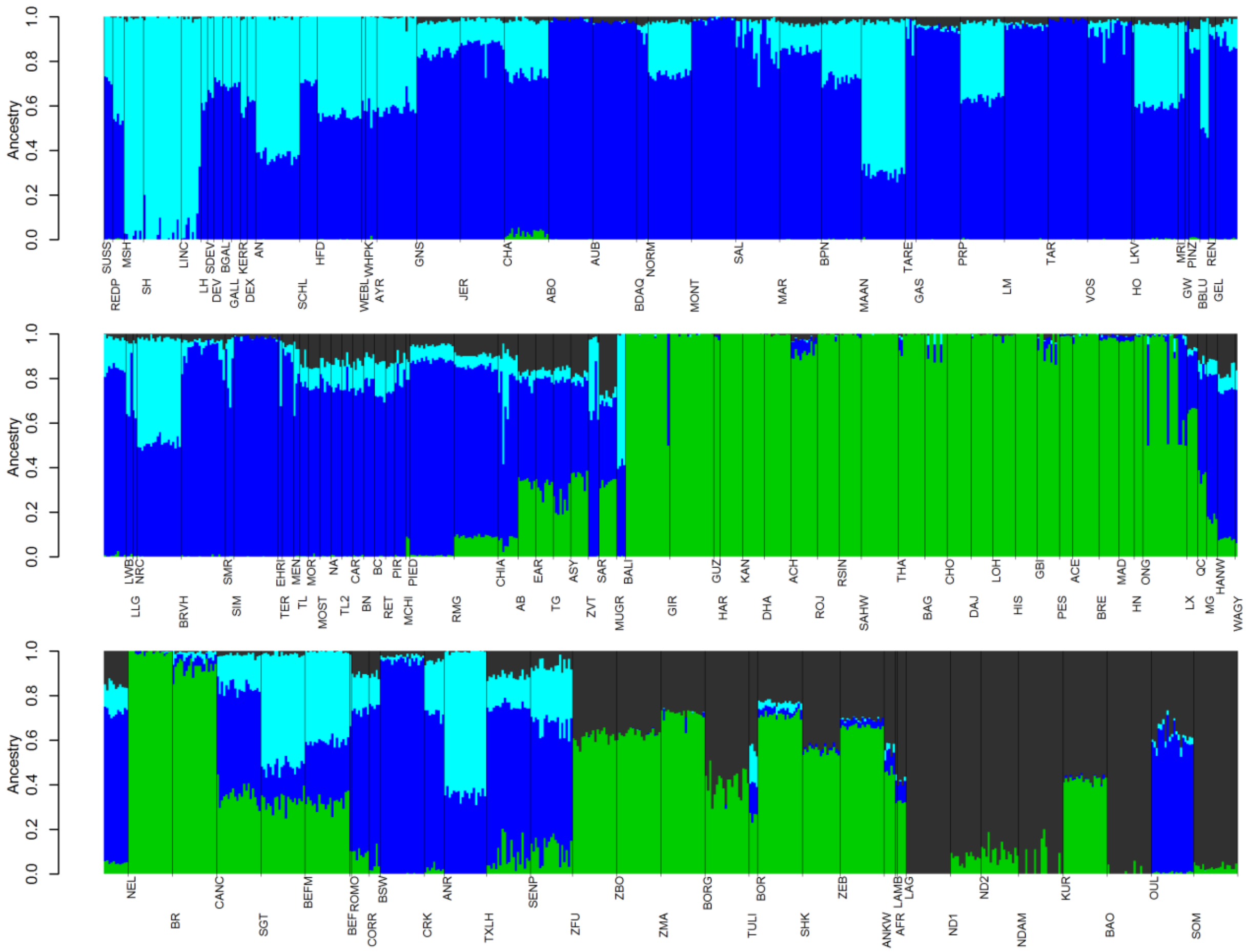

**Figure S6. Ancestry models with 4 ancestral populations (*K* = 4).** Blue represents Eurasian *Bos t. taurus* ancestry, green represents *Bos javanicus* and *Bos t. indicus* ancestry, dark grey represents African *Bos. t. taurus* ancestry, and cyan represents ancestry similar to Durham Shorthorns.

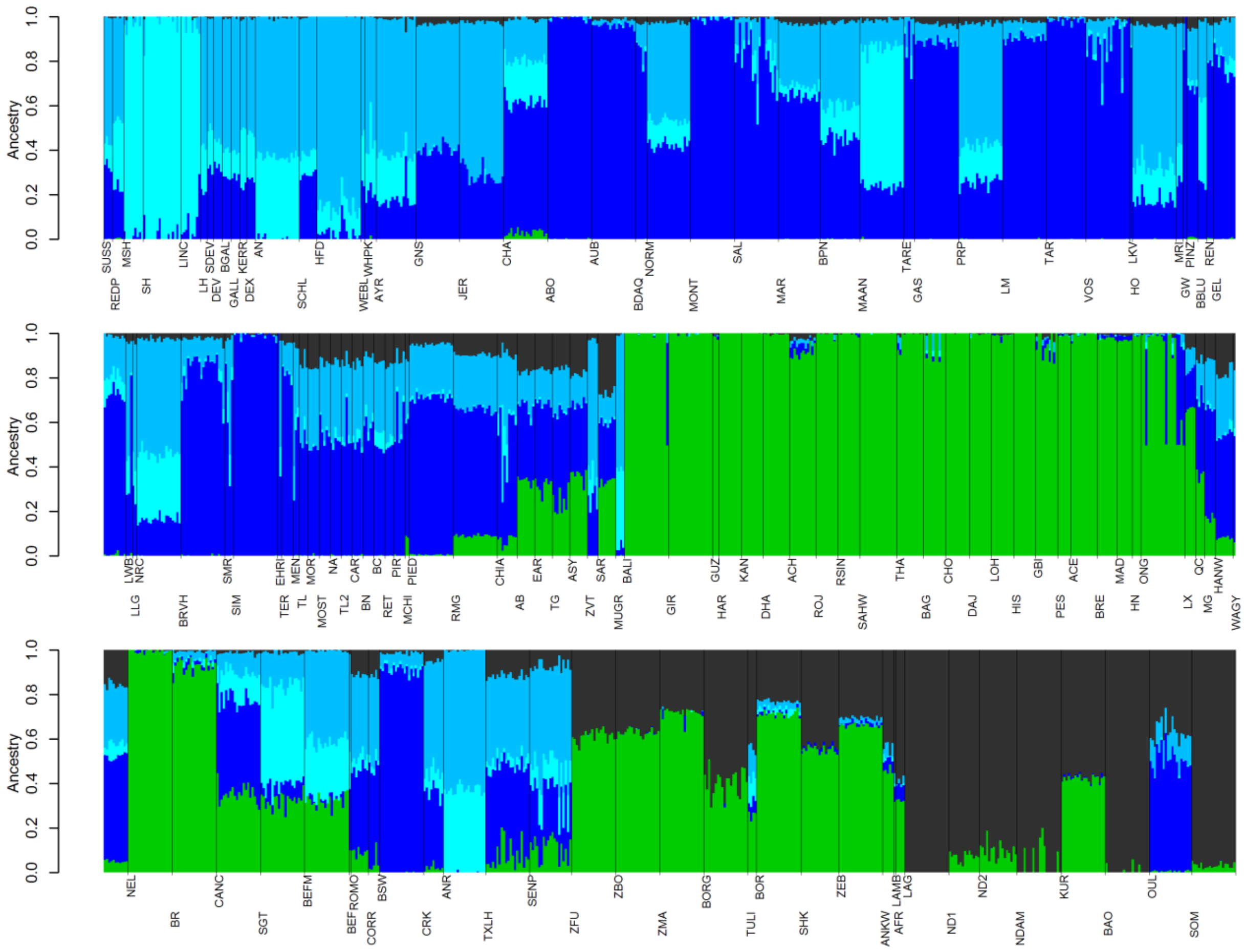

**Figure S7. Ancestry models with 5 ancestral populations (*K* = 5).** Blue represents Eurasian *Bos t. taurus* ancestry, green represents *Bos javanicus* and *Bos t. indicus* ancestry, dark grey represents African *Bos. t. taurus* ancestry, cyan represents ancestry similar to Durham Shorthorns, and deep sky blue represents British and Northern European ancestry.

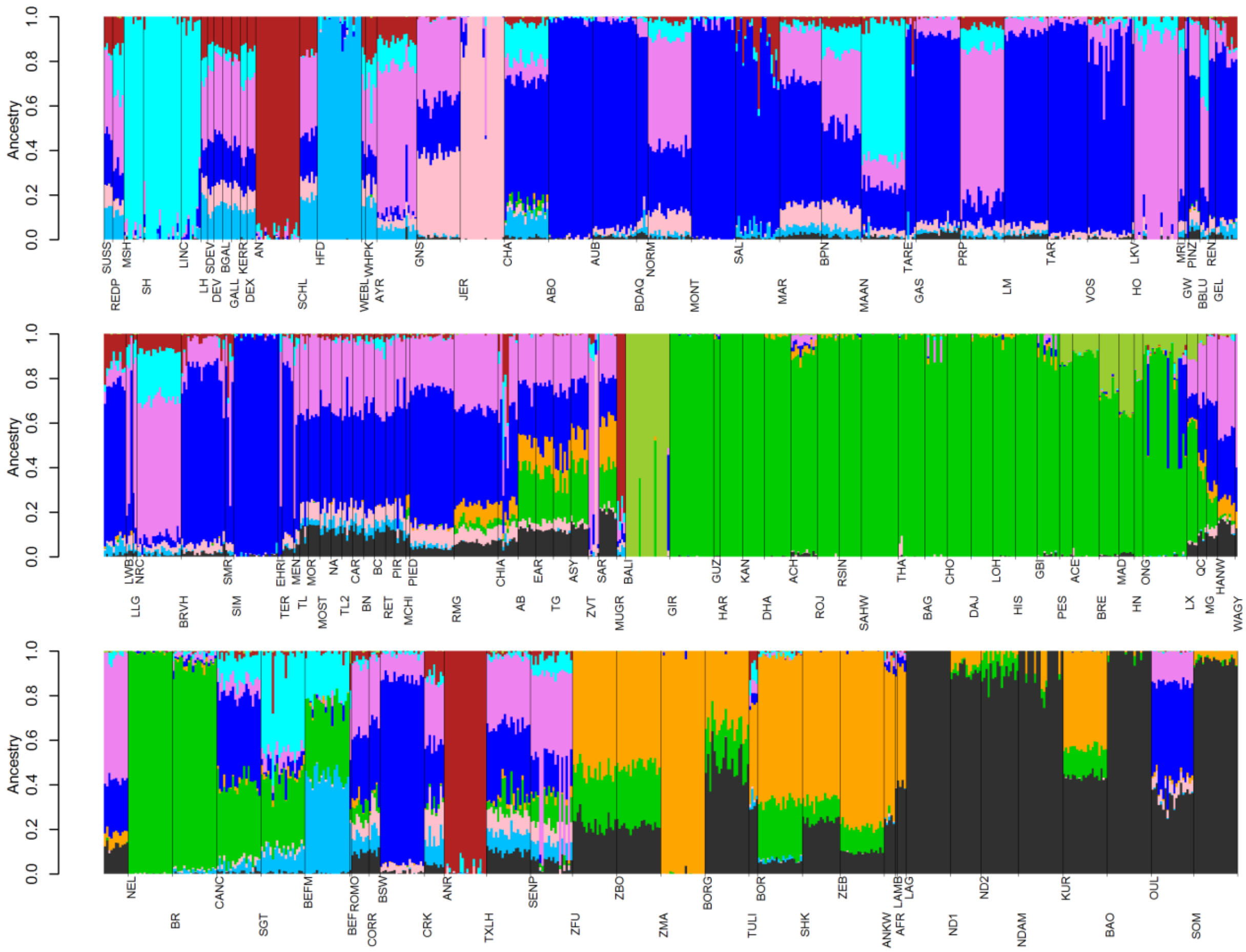

**Figure S8. Ancestry models with 10 ancestral populations (*K* = 10).**

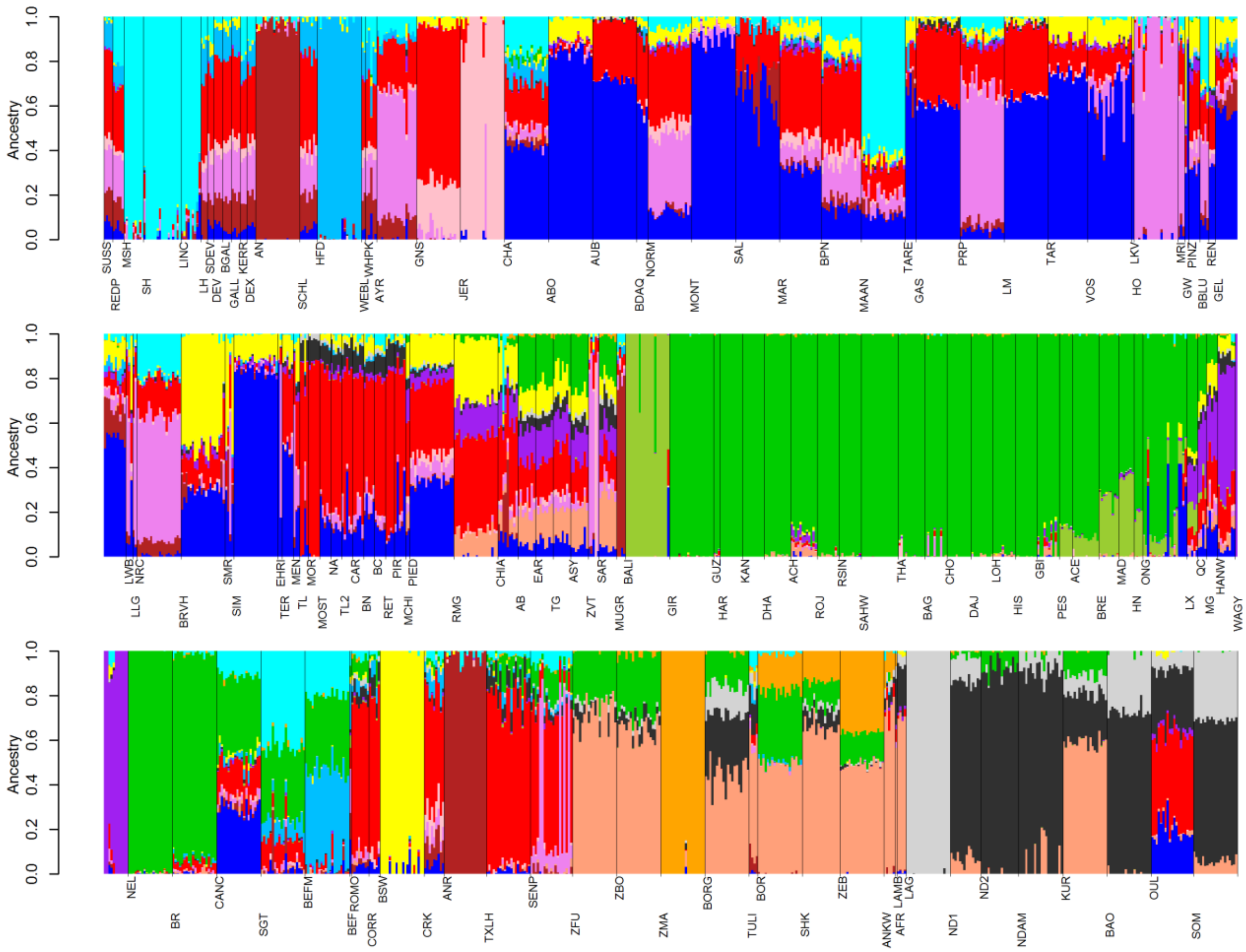

**Figure S9. Ancestry models with 15 ancestral populations (*K* = 15).**

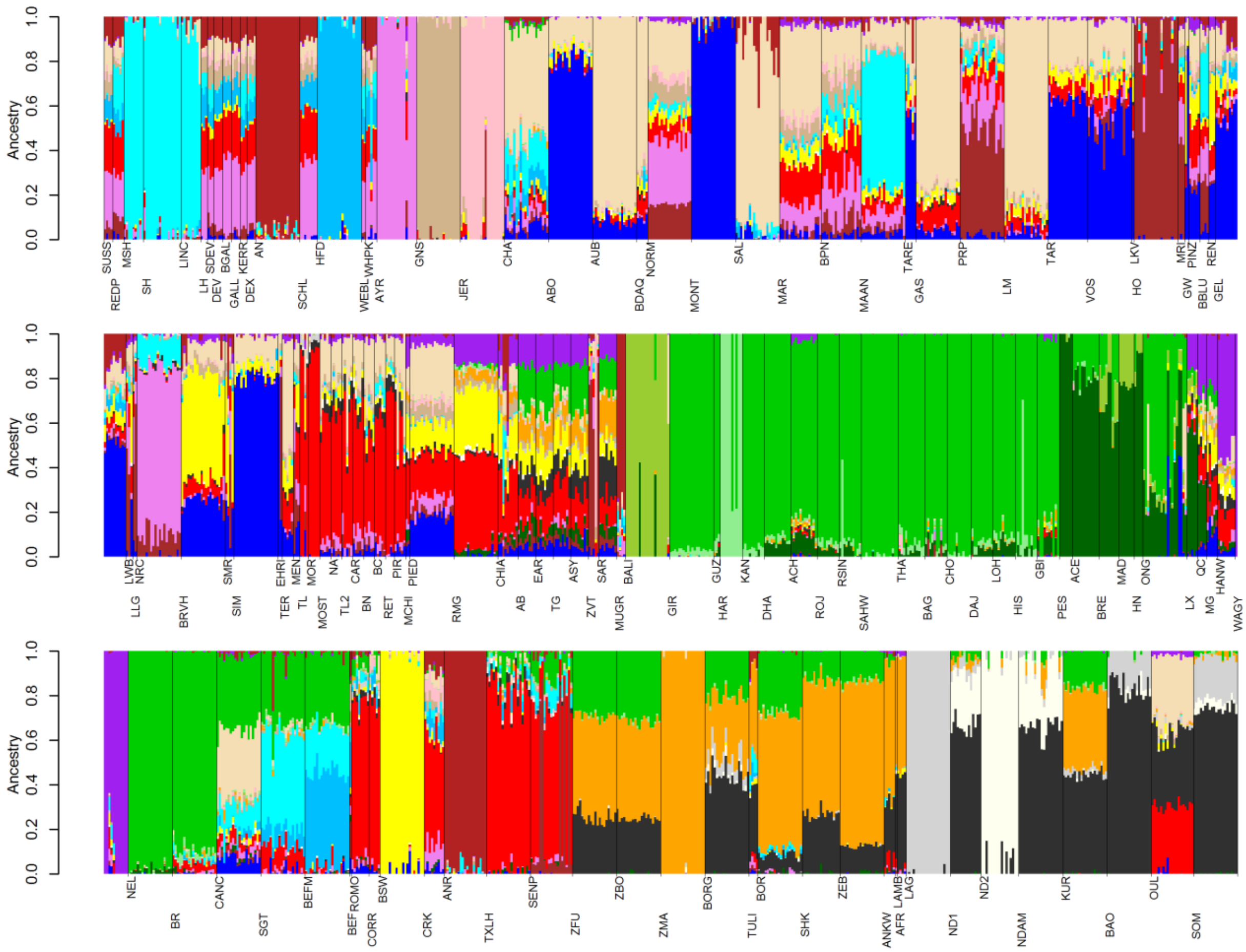

**Figure S10. Ancestry models with 20 ancestral populations (*K* = 20).**

## Tables

**Table 1. Significant $f_4$ statistics for African taurine breeds and populations[1].**

| Population A | Population B | Population C | Population D | $f_4$ | Standard Error | Z-score |
|---|---|---|---|---|---|---|
| N'Dama (ND2) | Shorthorn | Bali | Hariana | -0.00298 | 0.00061 | -4.91 |
| N'Dama (ND2) | Shorthorn | Bali | Sahiwal | -0.00254 | 0.00056 | -4.54 |
| N'Dama (ND2) | Montbeliard | Bali | Hariana | -0.00246 | 0.00051 | -4.82 |
| N'Dama (ND2) | Shorthorn | Bali | Gir | -0.00245 | 0.00058 | -4.21 |
| N'Dama (ND2) | Shorthorn | Bali | Aceh | -0.00217 | 0.00050 | -4.30 |
| N'Dama (ND2) | Shorthorn | Bali | Pesisir | -0.00206 | 0.00048 | -4.28 |
| N'Dama (ND2) | Montbeliard | Bali | Sahiwal | -0.00199 | 0.00048 | -4.11 |
| N'Dama (ND2) | Montbeliard | Bali | Gir | -0.00189 | 0.00053 | -3.55 |
| N'Dama (ND2) | Montbeliard | Bali | Aceh | -0.00175 | 0.00044 | -3.98 |
| ***N'Dama (NDAM)*** | ***Shorthorn*** | ***Bali*** | ***Hariana*** | ***-0.00156*** | ***0.00059*** | ***-2.67*** |
| N'Dama (ND2) | Montbeliard | Bali | Pesisir | -0.00151 | 0.00041 | -3.71 |
| Lagune | N'Dama | Hariana | Pesisir | -0.00136 | 0.00028 | -4.78 |
| ***Baoule*** | ***Shorthorn*** | ***Bali*** | ***Pesisir*** | ***-0.00134*** | ***0.00049*** | ***-2.73*** |
| Baoule | N'Dama (ND2) | Hariana | Pesisir | -0.00091 | 0.00028 | -3.18 |
| Lagune | N'Dama (ND2) | Hariana | Aceh | -0.00080 | 0.00024 | -3.35 |
| Lagune | N'Dama (ND2) | Hariana | Sahiwal | -0.00073 | 0.00019 | -3.84 |
| Lagune | N'Dama (ND2) | Gir | Pesisir | -0.00072 | 0.00023 | -3.10 |
| Baoule | N'Dama (ND2) | Gir | Pesisir | -0.00063 | 0.00019 | -3.31 |
| Lagune | N'Dama (ND2) | Pesisir | Aceh | 0.00055 | 0.00020 | 2.73 |
| ***N'Dama (NDAM)*** | ***Lagune*** | ***Hariana*** | ***Sahiwal*** | ***0.00056*** | ***0.00018*** | ***3.10*** |
| Lagune | N'Dama (ND2) | Gir | Hariana | 0.00064 | 0.00020 | 3.16 |
| Baoule | N'Dama (ND2) | Bali | Pesisir | 0.00072 | 0.00028 | 2.59 |
| ***N'Dama (NDAM)*** | ***Lagune*** | ***Hariana*** | ***Pesisir*** | ***0.00085*** | ***0.00022*** | ***3.81*** |
| N'Dama | N'Dama (ND2) | Bali | Pesisir | 0.00091 | 0.00025 | 3.62 |
| N'Dama | N'Dama (ND2) | Bali | Aceh | 0.00105 | 0.00026 | 4.09 |
| Baoule | N'Dama (ND2) | Bali | Aceh | 0.00112 | 0.00029 | 3.87 |
| N'Dama | N'Dama (ND2) | Bali | Gir | 0.00114 | 0.00028 | 4.08 |
| Baoule | N'Dama (ND2) | Bali | Sahiwal | 0.00122 | 0.00033 | 3.72 |
| N'Dama | N'Dama (ND2) | Bali | Sahiwal | 0.00125 | 0.00028 | 4.44 |
| Baoule | N'Dama (ND2) | Bali | Gir | 0.00135 | 0.00032 | 4.20 |
| Lagune | N'Dama (ND2) | Bali | Aceh | 0.00140 | 0.00033 | 4.23 |
| N'Dama | N'Dama (ND2) | Bali | Hariana | 0.00142 | 0.00031 | 4.55 |
| Lagune | N'Dama (ND2) | Bali | Sahiwal | 0.00148 | 0.00038 | 3.91 |
| Lagune | N'Dama (ND2) | Bali | Gir | 0.00157 | 0.00037 | 4.29 |
| Baoule | N'Dama (ND2) | Bali | Hariana | 0.00162 | 0.00036 | 4.47 |
| Lagune | N'Dama (ND2) | Bali | Hariana | 0.00221 | 0.00036 | 6.11 |

[1]Significant results with ND2 excluded from the analysis are indicated in bold italics.

**Table S1. Provenance for all samples included in the analyses.** Species and subspecies assignments are according to [32].

| Breed | Breed Code | No. Samples | No. from Decker *et al*. 2009 | No. from Gautier *et al*. 2010 | (Sub)Species | Continent | Geographic Origin |
|---|---|---|---|---|---|---|---|
| Bali | BALI | 20 | | | *Bos javanicus* | Asia | Bali, Indonesia |
| Zebu Fulani | ZFU | 20 | | 20 | *Bos t. indicus* | Africa | Benin |
| Zebu Bororo | ZBO | 20 | | 20 | *Bos t. indicus* | Africa | Chad |
| Zebu from Madagascar | ZMA | 20 | | 20 | *Bos t. indicus* | Africa | Madagascar |
| Nelore | NEL | 20 | 5 | | *Bos t. indicus* | Americas | Brazil |
| Brahman | BR | 20 | | | *Bos t. indicus* | Americas | Gulf Coast, United States |
| Ongole Grade | ONG | 20 | | | *Bos t. indicus* | Asia | Andhra Pradesh, India |
| Achai | ACH | 12 | | | *Bos t. indicus* | Asia | Khyber Pakhtun Khwa, Pakistan |
| Red Sindhi | RSIN | 10 | | | *Bos t. indicus* | Asia | Sindh, Pakistan |
| Cholistani | CHO | 11 | | | *Bos t. indicus* | Asia | Cholistan Desert, Punjab, Pakistan |
| Gir | GIR | 20 | 9 | | *Bos t. indicus* | Asia | Gujerat, India |
| Guzerat | GUZ | 3 | 3 | | *Bos t. indicus* | Asia | Guzarat, India |
| Hariana | HAR | 10 | | | *Bos t. indicus* | Asia | Haryana plains, India |
| Brebes | BRE | 9 | | | *Bos t. indicus* | Asia | Indonesia |
| Pesisir | PES | 6 | | | *Bos t. indicus* | Asia | Indonesia |
| Dajal | DAJ | 10 | | | *Bos t. indicus* | Asia | Punjab, Pakistan |
| Bhagnari | BAG | 10 | | | *Bos t. indicus* | Asia | Kaochi, Kalat, and Baluchistan, Pakistan |
| Gabrali | GBI | 10 | | | *Bos t. indicus* | Asia | Khyber Pakhtun Khwa, Pakistan |
| Hainan | HN | 4 | | | *Bos t. indicus* | Asia | Hainan Province, China |
| Madura | MAD | 7 | | | *Bos t. indicus* | Asia | Madura Island, Indonesia |
| Kankraj | KAN | 10 | | | *Bos t. indicus* | Asia | North Gujerat, India |
| Lohani | LOH | 10 | | | *Bos t. indicus* | Asia | Northwest Pakistan |
| Dhanni | DHA | 12 | | | *Bos t. indicus* | Asia | Punjab, Pakistan |
| Hissar | HIS | 10 | | | *Bos t. indicus* | Asia | Punjab, Pakistan |
| Sahiwal | SAHW | 17 | 10 | | *Bos t. indicus* | Asia | Punjab, Pakistan |
| Tharparkar | THA | 12 | | | *Bos t. indicus* | Asia | Southeast Sindh, Pakistan |
| Rojhan | ROJ | 10 | | | *Bos t. indicus* | Asia | Punjab, Pakistan |
| Aceh | ACE | 12 | | | *Bos t. indicus* | Asia | Sumatra, Indonesia |
| Lagune | LAG | 20 | | 20 | *Bos t. taurus* | Africa | Benin |
| Baoule | BAO | 20 | | 20 | *Bos t. taurus* | Africa | Burkina Faso |
| Kuri | KUR | 20 | | 20 | *Bos t. taurus* | Africa | Chad |
| N'Dama | NDAM | 20 | 4 | | *Bos t. taurus* | Africa | Ivory Coast, Africa |
| Oulmès Zaer | OUL | 19 | | 19 | *Bos t. taurus* | Africa | Morocco |
| N'Dama | ND1 | 14 | | 14 | *Bos t. taurus* | Africa | Southeast Burkina Faso |
| N'Dama | ND2 | 17 | | 17 | *Bos t. taurus* | Africa | Southwest Burkina Faso |
| Somba | SOM | 20 | | 20 | *Bos t. taurus* | Africa | Togo |
| Romosinuano | ROMO | 8 | 8 | | *Bos t. taurus* | Americas | Columbia |
| Florida Cracker | CRK | 9 | | | *Bos t. taurus* | Americas | Florida, United States |
| Corriente | CORR | 5 | 5 | | *Bos t. taurus* | Americas | Sonora, Mexico |
| Texas Longhorn | TXLH | 20 | 8 | | *Bos t. taurus* | Americas | Texas, United States |
| Brown Swiss | BSW | 20 | 7 | 8 | *Bos t. taurus* | Americas | United States |
| Red Angus | ANR | 19 | 6 | | *Bos t. taurus* | Americas | United States |
| Senepol | SENP | 19 | | | *Bos t. taurus* | Americas | United States Virgin Islands |
| Wagyu | WAGY | 12 | 6 | | *Bos t. taurus* | Asia | Japan |
| Hanwoo | HANW | 8 | | | *Bos t. taurus* | Asia | Korea |
| Mongolian | MG | 5 | | | *Bos t. taurus* | Asia | Mongolia1 |
| Murray Grey | MUGR | 4 | 4 | | *Bos t. taurus* | Australia | Australia |
| Angus | AN | 20 | 1 | | *Bos t. taurus* | Europe | Aberdeenshire, Scotland |
| Tarentaise | TARE | 5 | 5 | | *Bos t. taurus* | Europe | Alpine Massif-Central of southeastern France |
| Pinzgauer | PINZ | 5 | 5 | | *Bos t. taurus* | Europe | Austria |
| Belgian Blue | BBLU | 4 | 4 | | *Bos t. taurus* | Europe | Belgium |
| Simmental | SIM | 20 | | | *Bos t. taurus* | Europe | Bern, Switzerland |
| Simmentaler | SMR | 4 | | | *Bos t. taurus* | Europe | Bern, Switzerland |
| Maine-Anjou | MAAN | 20 | 5 | 15 | *Bos t. taurus* | Europe | Brittany, France |
| Rendena | REN | 3 | | | *Bos t. taurus* | Europe | Central Alps |
| Gelbvieh | GEL | 20 | 5 | | *Bos t. taurus* | Europe | Central Germany |
| Berrenda en Negro | BN | 5 | | | *Bos t. taurus* | Europe | Ciudad Real, Jaen, Cordoba, Sevilla, and Huelva, Spain |
| Berrenda en Colorado | BC | 5 | | | *Bos t. taurus* | Europe | Cordoba, Sevilla, Huelva, and Cadiz, Spain |
| Devon | DEV | 4 | 4 | | *Bos t. taurus* | Europe | Devon, England |
| Romagnola | RMG | 20 | 10 | | *Bos t. taurus* | Europe | Emilia, Italy |
| Beef Shorthorn | SH | 17 | 7 | | *Bos t. taurus* | Europe | England |
| Lincoln Red | LINC | 9 | 9 | | *Bos t. taurus* | Europe | England |
| Milking Shorthorn | MSH | 9 | 1 | | *Bos t. taurus* | Europe | England |
| Montbeliard | MONT | 20 | 2 | 18 | *Bos t. taurus* | Europe | France |
| Normande | NORM | 20 | 1 | 19 | *Bos t. taurus* | Europe | France |
| Guernsey | GNS | 20 | 10 | | *Bos t. taurus* | Europe | Guernsey Island |
| Holstein | HO | 20 | | | *Bos t. taurus* | Europe | Holland |
| Dexter | DEX | 4 | 4 | | *Bos t. taurus* | Europe | Ireland |
| Kerry | KERR | 3 | 3 | | *Bos t. taurus* | Europe | Ireland |
| Marchigiana | MCHI | 2 | 2 | | *Bos t. taurus* | Europe | Italy |
| Jersey | JER | 20 | 7 | | *Bos t. taurus* | Europe | Jersey Island |
| Lithuanian Light Grey | LLG | 2 | | | *Bos t. taurus* | Europe | Lithuania |
| Lithuanian White Backed | LWB | 3 | | | *Bos t. taurus* | Europe | Lithuania |
| Limousin | LM | 20 | | | *Bos t. taurus* | Europe | Massif Central, France |
| Salers | SAL | 20 | 4 | | *Bos t. taurus* | Europe | Massif Central, France |
| Menorquina | MEN | 3 | | | *Bos t. taurus* | Europe | Menorca, Spain |
| Mostrenca | MOST | 5 | | | *Bos t. taurus* | Europe | National Park of Donana, southwestern Spain |
| Groningen Whitehead | GW | 2 | | | *Bos t. taurus* | Europe | Netherlands |
| Lakenvelder | LKV | 1 | | | *Bos t. taurus* | Europe | Netherlands |
| Meuse-Rhine-Ijjsel | MRI | 3 | | | *Bos t. taurus* | Europe | Netherlands |
| Red Poll | REDP | 5 | 5 | | *Bos t. taurus* | Europe | Norfolk and Suffolk, England |
| Vosgienne | VOS | 20 | | 20 | *Bos t. taurus* | Europe | Northeast France |
| Bretonne Black Pied | BPN | 18 | | 18 | *Bos t. taurus* | Europe | Northwest France |
| French Red Pied Lowland | PRP | 20 | | 20 | *Bos t. taurus* | Europe | Northwest France |
| Maraichine (Parthenaise) | MAR | 19 | | 19 | *Bos t. taurus* | Europe | Northwest France |
| Piedmontese | PIED | 20 | 9 | | *Bos t. taurus* | Europe | Northwest Italy |
| Pirenaica | PIR | 5 | | | *Bos t. taurus* | Europe | northwest of Spain |
| Norwegian Red | NRC | 20 | 9 | | *Bos t. taurus* | Europe | Norway |
| Blonde d'Aquitaine | BDAQ | 5 | 5 | | *Bos t. taurus* | Europe | Pyrenees, France |
| Morucha | MOR | 5 | | | *Bos t. taurus* | Europe | Salamanca |
| Charolais | CHA | 20 | | | *Bos t. taurus* | Europe | Saône-et-Loire, France |
| Belted Galloway | BGAL | 4 | 4 | | *Bos t. taurus* | Europe | Scotland |
| Galloway | GALL | 4 | 4 | | *Bos t. taurus* | Europe | Scotland |
| Finnish Ayrshire | AYR | 18 | 2 | | *Bos t. taurus* | Europe | Scotland/Finland |
| Negra Andaluza | NA | 5 | | | *Bos t. taurus* | Europe | Sierra Morena Mountains, Cordoba, and Sevilla Spain |
| Cardena Andaluza | CAR | 5 | | | *Bos t. taurus* | Europe | Sierra Morena, Spain |
| South Devon | SDEV | 3 | 3 | | *Bos t. taurus* | Europe | South England |
| Aubrac | AUB | 20 | | 20 | *Bos t. taurus* | Europe | South France |
| Sussex | SUSS | 4 | 4 | | *Bos t. taurus* | Europe | Southeast England |
| Abondance | ABO | 20 | | 20 | *Bos t. taurus* | Europe | Southeast France |
| Tarine | TAR | 18 | | 18 | *Bos t. taurus* | Europe | Southeast France |
| Gascon | GAS | 20 | | 20 | *Bos t. taurus* | Europe | Southwest France |
| Retinta | RET | 4 | | | *Bos t. taurus* | Europe | Southwest of Spain and bordering Portugal |
| Toro de Lidia | TL | 4 | | | *Bos t. taurus* | Europe | Spain |
| Toro de Lidia | TL2 | 5 | | | *Bos t. taurus* | Europe | Spain |
| Braunvieh | BRVH | 20 | | | *Bos t. taurus* | Europe | Switzerland |
| Ehringer | EHRI | 2 | | | *Bos t. taurus* | Europe | Switzerland |
| Anatolian Black | AB | 8 | | | *Bos t. taurus* | Europe | Turkey |
| Anatolian Southern Yellow | ASY | 8 | | | *Bos t. taurus* | Europe | Turkey |
| East Anatolian Red | EAR | 8 | | | *Bos t. taurus* | Europe | Turkey |
| South Anatolian Red | SAR | 8 | | | *Bos t. taurus* | Europe | Turkey |
| Turkish Grey | TG | 8 | | | *Bos t. taurus* | Europe | Turkey |
| Zavot | ZVT | 5 | | | *Bos t. taurus* | Europe | Turkey |
| Terrana | TER | 5 | | | *Bos t. taurus* | Europe | Vasconcades mountainous region of Alava, Spain |
| Hereford | HFD | 20 | 1 | | *Bos t. taurus* | Europe | Wales |
| Welsh Black | WEBL | 2 | 2 | | *Bos t. taurus* | Europe | Wales |
| White Park | WHPK | 5 | 4 | | *Bos t. taurus* | Europe | Wales |
| Chianina | CHIA | 9 | 7 | | *Bos t. taurus* | Europe | West Central Italy |
| Scottish Highland | SCHL | 8 | 8 | | *Bos t. taurus* | Europe | Western Scottland |
| Longhorn | LH | 3 | 3 | | *Bos t. taurus* | Europe | Yorkshire, England |
| Borgou | BORG | 20 | | 20 | Hybrid | Africa | Benin |
| Tuli | TULI | 4 | | | Hybrid | Africa | Botswana |
| Landim | LAMB | 1 | | | Hybrid | Africa | East Coast of South Africa |
| Sheko | SHK | 17 | | | Hybrid | Africa | Ethiopia |
| East African Shorthorn Zebu | ZEB | 20 | | | Hybrid | Africa | Kenya |
| Ankole-Watusi | ANKW | 5 | | | Hybrid | Africa | Ruanda |
| Africander | AFR | 4 | | | Hybrid | Africa | South Africa |
| Boran | BOR | 20 | | | Hybrid | Africa | southern Ethiopia |
| Canchim | CANC | 20 | | | Hybrid | Americas | Brazil |
| Beefalo | BEF | 1 | | | Hybrid | Americas | Northwest United States |
| Beefmaster | BEFM | 20 | | | Hybrid | Americas | Texas, United States |
| Santa Gertrudis | SGT | 20 | | | Hybrid | Americas | Texas, United States |
| Luxi | LX | 5 | | | Hybrid | Asia | Shandong Province, China |
| Qinchuan | QC | 4 | | | Hybrid | Asia | Shaanxi Province, China |

**Table S2. Cross-validation and *ΔK* values for ADMIXTURE ancestry models with *K* ranging from 1 to 20.**

| *K* | Cross-validation error | Log-likelihood | *L*'(*K*) | *L*''(*K*) |
|---|---|---|---|---|
| 1 | 0.63636 | -65702704 | N/A | N/A |
| 2 | 0.54374 | -59538328 | 6164375.96 | 4545316.79 |
| 3 | 0.51985 | -57919269 | 1619059.17 | 1105462.91 |
| 4 | 0.51288 | -57405673 | 513596.27 | 253401.84 |
| 5 | 0.50983 | -57145478 | 260194.42 | 26732.82 |
| 6 | 0.50693 | -56912017 | 233461.61 | 9346.55 |
| 7 | 0.50424 | -56687902 | 224115.05 | -163722.93 |
| 8 | 0.49888 | -56300064 | 387837.99 | 450205.28 |
| 9 | 0.50094 | -56362431 | -62367.29 | -547943.87 |
| 10 | 0.49428 | -55876855 | 485576.58 | 382526.32 |
| 11 | 0.49325 | -55773804 | 103050.26 | -68045.84 |
| 12 | 0.49175 | -55602708 | 171096.10 | 46476.86 |
| 13 | 0.49041 | -55478089 | 124619.24 | -79680.06 |
| 14 | 0.48836 | -55273790 | 204299.30 | 81316.25 |
| 15 | 0.48708 | -55150807 | 122983.05 | 37711.26 |
| 16 | 0.48640 | -55065535 | 85271.79 | -65609.06 |
| 17 | 0.48497 | -54914654 | 150880.84 | 72573.21 |
| 18 | 0.48455 | -54836346 | 78307.63 | -63149.16 |
| 19 | 0.48293 | -54694890 | 141456.79 | 38935.55 |
| 20 | 0.48206 | -54592368 | 102521.24 | 102521.24 |

**Table S3. Five most negative and significant $f_3$ statistics for Brebes and Madura showing Bali (*Bos javanicus*) introgression.**

| Population A | Population B | Population C | $f_3$ | Standard Error | Z-score |
|---|---|---|---|---|---|
| Brebes | Bali | Lohani | -0.00765 | 0.00041 | -18.80 |
| Brebes | Nelore | Bali | -0.00769 | 0.00041 | -18.76 |
| Brebes | Red Sindhi | Bali | -0.00747 | 0.00040 | -18.73 |
| Brebes | Brahman | Bali | -0.00708 | 0.00039 | -18.24 |
| Brebes | Sahiwal | Bali | -0.00764 | 0.00042 | -18.07 |
| | | | | | |
| Madura | Bali | Lohani | -0.00656 | 0.00047 | -13.96 |
| Madura | Aceh | Bali | -0.00661 | 0.00048 | -13.86 |
| Madura | Rojhan | Bali | -0.00605 | 0.00046 | -13.09 |
| Madura | Achai | Bali | -0.00601 | 0.00046 | -13.08 |
| Madura | Sahiwal | Bali | -0.00639 | 0.00049 | -13.07 |

**Table S4. Five most negative and significant $f_3$ statistics for Maine-Anjou, Santa Gertrudis, and Beefmaster showing Shorthorn admixture.**

| Population A | Population B | Population C | $f_3$ | Standard Error | Z-score |
|---|---|---|---|---|---|
| Maine Anjou | Tarine | Beef Shorthorn | -0.00389 | 0.00057 | -6.84 |
| Maine Anjou | Aubrac | Beef Shorthorn | -0.00359 | 0.00056 | -6.47 |
| Maine Anjou | Tarine | Milking Shorthorn | -0.00391 | 0.00061 | -6.36 |
| Maine Anjou | Africander | Beef Shorthorn | -0.00357 | 0.00057 | -6.30 |
| Maine Anjou | Somba | Beef Shorthorn | -0.00376 | 0.00060 | -6.29 |
| | | | | | |
| Santa Gertrudis | Tharparkar | Beef Shorthorn | -0.02641 | 0.00071 | -37.12 |
| Santa Gertrudis | Rojhan | Beef Shorthorn | -0.02664 | 0.00072 | -37.11 |
| Santa Gertrudis | Aceh | Beef Shorthorn | -0.02618 | 0.00071 | -36.80 |
| Santa Gertrudis | Dajal | Beef Shorthorn | -0.02674 | 0.00073 | -36.50 |
| Santa Gertrudis | Kankraj | Beef Shorthorn | -0.02681 | 0.00074 | -36.47 |
| | | | | | |
| Beefmaster | Gir | Beef Shorthorn | -0.02004 | 0.00099 | -20.22 |
| Beefmaster | Beef Shorthorn | Guzerat | -0.01943 | 0.00099 | -19.60 |
| Beefmaster | Dajal | Beef Shorthorn | -0.01965 | 0.00101 | -19.54 |
| Beefmaster | Beef Shorthorn | Dhanni | -0.01965 | 0.00101 | -19.52 |
| Beefmaster | Tharparkar | Beef Shorthorn | -0.01933 | 0.00100 | -19.26 |